\documentclass[journal,12pt,draftclsnofoot,onecolumn,english]{IEEEtran}

\usepackage{epsfig}
\usepackage{times}
\usepackage{float}
\usepackage{afterpage}
\usepackage{amsmath}
\usepackage{amstext,cite}
\usepackage{amssymb,bm}
\usepackage{latexsym}
\usepackage{color}
\usepackage{graphicx}
\usepackage{amsmath}
\usepackage{amsthm}
\usepackage{graphicx}
\usepackage[center]{caption}
\usepackage{pstricks}
\usepackage{caption}
\usepackage{subcaption}
\usepackage{booktabs}
\usepackage{multicol}
\usepackage{lipsum}
\usepackage[T1]{fontenc}
\usepackage{hyperref}
 \usepackage{aecompl}
 \usepackage{mathrsfs}

 \usepackage{array}
\usepackage{multirow}

\providecommand{\tabularnewline}{\\}

 \usepackage{babel}

\allowdisplaybreaks

\long\def\comment#1{}

\newcommand{\dv}{{\mathbf d}}

\newcommand{\fv}{{\mathbf f}}
\newcommand{\gv}{{\mathbf g}}

\newcommand{\pv}{{\mathbf p}}
\newcommand{\qv}{{\mathbf q}}

\newcommand{\Cc}{{\mathcal C}}

\newcommand{\Hc}{{\mathcal H}}

\newcommand{\Oc}{{\mathcal O}}
\newcommand{\Pc}{{\mathcal P}}
\newcommand{\Qc}{{\mathcal Q}}

\newcommand{\Sc}{{\mathcal S}}

\newcommand{\Wc}{{\mathcal W}}

\newcommand{\Bsf}{{\mathsf B}}

\newcommand{\Ksf}{{\mathsf K}}
\newcommand{\Lsf}{{\mathsf L}}
\newcommand{\Msf}{{\mathsf M}}
\newcommand{\Nsf}{{\mathsf N}}

\newcommand{\Rsf}{{\mathsf R}}

\newcommand{\Usf}{{\mathsf U}}

\newtheorem{thm}{Theorem}

\newtheorem{lem}{Lemma}

\newtheorem{rem}{Remark}
\newtheorem{example}{Example}

\providecommand{\definitionname}{Definition}

\providecommand{\tabularnewline}{\\}

\begin{document}

\title{On Coded Caching with Private Demands}

\author{
Kai~Wan,~\IEEEmembership{Member,~IEEE,} 
and~Giuseppe Caire,~\IEEEmembership{Fellow,~IEEE}
\thanks{
K.~Wan and G.~Caire are with the Electrical Engineering and Computer
Science Department, Technische Universit\"at Berlin, 10587 Berlin, Germany
(e-mail:  kai.wan@tu-berlin.de; caire@tu-berlin.de). The work of K.~Wan and G.~Caire was partially funded by the European Research Council  under the ERC Advanced Grant N. 789190, CARENET.}
}
\maketitle

\begin{abstract}
 Caching is an efficient way to reduce network traffic congestion during peak hours by storing some
content at the user's local cache memory without knowledge of later demands. For the 
shared-link caching model, Maddah-Ali and Niesen (MAN) proposed a two-phase  ({\it placement} and {\it delivery}) coded caching strategy, which  is order optimal within a constant factor. However, in 
the MAN coded caching scheme, each user can obtain the information about the demands of other users, i.e.,
the MAN coded caching scheme is inherently prone to tampering and spying the activity/demands of other users.
In this paper, we formulate an information-theoretic shared-link    caching model with private demands, where there are $\Ksf$ cache-aided users (which can cache up to $\Msf$ files) connected to a central server with access to $\Nsf$ files. Each user requests $\Lsf$ files. 
 Our objective is to design a two-phase private caching  scheme with minimum load while preserving the information-theoretic privacy of the demands of each user with respect to other users. 

A trivial solution is the uncoded caching scheme  which lets each user recover all the $\Nsf$ files, referred to as {\it baseline scheme}.   For this problem we propose two novel schemes which achieve the information-theoretic privacy of the users' demands while also achieving a non-trivial caching gain over the baseline scheme. The general underlying idea is to satisfy the users' requests by generating a set of coded multicast messages that is symmetric with respect to the library files, such that for each user $k$,  the mutual information between these messages and the   demands of all other users given the cache content and the demand of user $k$ is zero. 
In the first scheme, referred to as virtual-user scheme, we introduce a number of virtual users such that each $\Lsf$-subset of files is demanded by $\Ksf$ real or virtual (effective) 
users and use the MAN delivery to generate multicast messages. From the viewpoint of each user, the set of multicast
messages is symmetric over all files even if each single multicast message is not.
This scheme incurs in an extremely large sub-packetization. Then, we propose a second scheme, referred to as MDS-based scheme,  based on a novel MDS-coded cache placement. In this case, we generate multicast messages where each multicast message contains one MDS-coded symbol from each file in the library and thus is again symmetric over all the files from the viewpoint of each user. The sub-packetization level of the MDS-based scheme is exponentially smaller than that needed by the virtual-user scheme.  
 
Compared  with the   existing shared-link coded caching converse bounds  without privacy, the virtual-user scheme is proved to be order optimal with a constant factor when $\Nsf \leq \Lsf\Ksf$,  or when $\Nsf\geq \Lsf\Ksf$ and $\Msf\geq \Nsf/\Ksf$. In addition, when $\Msf \geq \Nsf/2$, both of the virtual-user scheme and the MDS-based scheme are order optimal within a factor of $2$.

\end{abstract}



\section{Introduction}
\label{sec:intro}
\subsection{Brief Review of Coded Caching}  
\label{sub:Brief Description}
Recent years have witnessed a steep increase of wireless devices connected to the Internet, leading to a heavy  network traffic because of multimedia streaming, web-browsing   and
social networking.  Furthermore,   the high temporal
variability of network traffic  results in congestions during peak-traffic times and
underutilization of the network during off-peak times.
{\it Caching} is a promising technique  to  
 reduce  peak traffic by
taking advantage of memories distributed across the network to duplicate content during
off-peak times~\cite{5gcaching} . 
With the help of caching, network traffic could be shifted  from peak to  off-peak
hours in order to  smooth out the traffic load and reduce congestion.
In the seminal paper~\cite{dvbt2fundamental}, an information-theoretic and network-coding theoretic model for caching was proposed. In this model, 
two phases are included in a caching system:
i) {\it placement phase}: each user equipped  with cache stores  some bits in its cache component without knowledge of later demands;
ii) {\it delivery phase}: after each user has made its request  and according to cache contents, the server transmits packets such that  each user can recover its desired file(s). The goal is to minimize the number of  transmitted bits (referred to as {\it load} in this paper).

Coded caching strategy was originally proposed in~\cite{dvbt2fundamental} for the  shared-link broadcast networks where a server with  a library of $\Nsf$ files, of $\Bsf$ bits each, is connected to $\Ksf$ users (each of which is with a cache of $\Msf\Bsf$ bits) through a shared error-free broadcast link. Each user requests one file independently in the delivery phase. 
 Maddah-Ali and Niesen (MAN) proposed a    coded caching scheme  that utilizes an uncoded combinatorial cache construction in the placement phase and a binary linear network code to generate multicast messages in the delivery phase.  
For $\Msf = t \frac{\Nsf}{\Ksf}$ with $t\in[0:\Ksf],$ 
the transmitted
load is $\frac{\Ksf (1-\Msf/\Nsf)}{1+\Ksf\Msf/\Nsf}$. For other memory size, the lower convex envelope of the above memory-load tradeoff points is achievable by memory-sharing between schemes for integer values of $t=\Ksf\Msf/\Nsf$.  
Compared to the conventional uncoded caching scheme which lets each user store $\Msf \Bsf/\Nsf$ bits of each file in the placement phase and broadcasts the uncached part of each desired file   during the delivery phase with the transmitted load $\Ksf (1-\Msf/\Nsf)$, the MAN coded caching scheme has an additional coded caching gain (i.e., load reduction factor) equal to $1+\Ksf\Msf/\Nsf$. 
It was proved in~\cite{ontheoptimality} that the worst-case load achieved by the MAN coded caching scheme among all possible demands is optimal under the constraint of uncoded placement (i.e., each user directly stores a subset of bits in the library) and $\Nsf \geq \Ksf$.  When $\Nsf\geq \Ksf$, the MAN coded caching scheme was also proved in~\cite{yas2} to be generally order optimal within a factor of $2$. For any $\Nsf$ and $\Ksf$, 
a factor of $4$ for the order optimality of the MAN coded caching scheme was proved  in~\cite{improvedlower2017Ghasemi}.
 By observing that some MAN multicast messages  are redundant for the case $\Nsf<\Ksf$, the authors in~\cite{exactrateuncoded} proposed an improved delivery scheme which is optimal under the constraint of uncoded cache placement for any $\Nsf$ and $\Ksf$, and optimal within a factor of $2$ over all possible placement strategies.

In the MAN caching model, each user requests only one file, which may not be practical. 
The caching problem with   multi-request was originally considered in~\cite{ji2015multirequest} where  each user demands $\Lsf$ files from the library. 
With the MAN placement,   to   divide   the delivery phase into $\Lsf$ rounds where in each round the MAN coded caching scheme in~\cite{dvbt2fundamental} (referred to as {\it $\Lsf$-round MAN coded caching scheme} in this paper, which reduces to the MAN coded caching scheme when $\Lsf=1$) is used to let each user decode one file, can achieve a generally order optimal worst-case load   within a factor  of $18$~\cite{ji2015multirequest}. By further tightening the converse bound, this order optimality factor was reduced to $11$ in~\cite{Sengupta2017multirequest}.

The MAN coded caching strategy was also used in a number of extended models, such as  decentralized
caching where users must fill their caches independently of   other users~\cite{decentralizedcoded}, device-to-device (D2D) caching systems where users communicate among each other during the delivery phase~\cite{d2dcaching}, cache-aided topological networks where the server communicates with the users through some intermediate relays~\cite{multiserver,cachingJi2015,wan2017novelmulticase}, etc.  However,  these extended models will not be considered in our paper, and thus we do not go into details.  

\subsection{Existing Secure Coded Caching Schemes}  
\label{sub:exsiting privacy}
Soon after the appearance of~\cite{dvbt2fundamental}, various `secure' versions of the caching problem   have been proposed.
Secure coded caching was originally considered in~\cite{securedelivery2015Avik}, where there are some wiretappers who can also receive the broadcasted packets from the server. To prevent the wiretappers from obtaining any information about the files in the library,  the authors in~\cite{securedelivery2015Avik} let each user store not only the content about the library in its cache, but also some `keys'. In the delivery phase, each multicast message is generated by taking  XOR of the MAN multicast message and some key   in order to `lock'  the multicast messages such that only the intended users   can unlock it. This secure caching scheme against wiretappers was proved in~\cite{towardssecu2017Bahrami} to be optimal under the constraint of uncoded cache placement.
Another secure shared-link caching model was proposed in~\cite{privatecaching2018Vai}. In this case, the objective is to  avoid each user to get any information about the files not required by that user. The placement and delivery phases were designed based on the MAN coded caching scheme with an additional
secret sharing precoding~\cite{share1979Shamir} on each file (i.e., by encoding a message with $(n,t)$ secret sharing code where $n>t$,    any $t$ shares do
not reveal any information about the message and the message can be reconstructed
from all the $n$ shares). In addition, the secure caching scheme in~\cite{privatecaching2018Vai} could also successfully prevent external wiretappers, because each multicast message is also  locked  by a  key.
The above strategies to prevent external wiretappers and internal malicious users from retrieving information about the library, were then used in extended models, such as D2D caching systems~\cite{d2dsecu2018Zewail,boundond2d2018Awan}, topological cache-aided relay networks~\cite{combinationsecu2018Zewail}, erasure broadcast channels~\cite{erasuresecu2018Kamel}, etc.

\subsection{Coded Caching with Private Demands}  
\label{sub:private demand}

The existing secure caching schemes are based on the MAN coded caching scheme (with or without a secure precoding on each file) and then generate  locked     MAN multicast  messages. 
 However, a malicious user could simply use the MAN multicast messages (or  locked    MAN multicast  messages) in order to learn the requests of other users, 
 e.g., to perform some survey on user preferences, which is not good in terms of privacy.  
Shared-link caching problem with single request to preserve the users demands from other users was originally considered in~\cite{Engle2017privatecaching}. The caching scheme proposed in~\cite{Engle2017privatecaching} generates $\ell$ virtual users each of which randomly demands one file, such that each user cannot match the exact request to any other user. However, this caching scheme is not completely private from an information-theoretic viewpoint. For example, if there exists undemanded file by any real or virtual user, each user will know this file has not been demanded such that it can get some information about the users demands from the transmission.
 In this paper, we formulate an information-theoretic caching problem which aims to preserve the privacy of  the demands of each user 
with respect to other users during the transmission.

Let us focus on a toy example with $\Ksf=2$, $\Nsf=3$ and $\Msf=2\Nsf/3=2$.  In this example, $t=\Ksf\Msf/\Nsf=4/3$, which is not an integer, and thus we should use the memory-sharing between $\Msf_1=\Nsf t_1/\Ksf=3/2$ with $t_1=1$ and  $\Msf_2=\Nsf t_2/\Ksf=3$ with $t_2=2$.
By the MAN placement, we divide each file into three equal-length and  non-overlapping subfiles, the $i^{\text{th}}$ file, denoted by $F_i$, has three subfiles $F_{i,\{1\}}$, $F_{i,\{2\}}$, and $F_{i,\{1,2\}}$. User $1$ caches $F_{i,\{1\}}$ and $F_{i,\{1,2\}}$, while user $2$ caches $F_{i,\{2\}}$ and $F_{i,\{1,2\}}$. 

In the delivery phase, we consider two demands:
\begin{itemize}
\item if the demand is $(1,2)$, i.e., user $1$ demands file $F_1$ and user $2$ demands file $F_2$, we transmit the MAN multicast message $F_{1,\{2\}}\oplus F_{2,\{1\}}$, where $\oplus$ represents the XOR operation, such that user $1$ can recover $F_{1,\{2\}}$  and user $2$ can recover $F_{2,\{1\}}$. However, for the sake of successful decoding, user $1$ needs to know that $ F_{2,\{1\}}$ is contained by the multicast message, and thus it knows user $2$ demands $F_2$. Similarly,  user $2$ will know the demand of user $1$.
\item if the demand is $(1,1)$, i.e., both users $1$ and $2$ demand  file $F_1$, we transmit the MAN multicast message $F_{1,\{2\}}\oplus F_{1,\{1\}}$, such that user $1$ can recover $F_{1,\{2\}}$  and user $2$ can recover $F_{1,\{1\}}$. However, from the transmission,  user $1$   knows user $2$ demands $F_1$, while user $2$ knows the demand of user $1$.
\end{itemize}


The above example shows that the MAN scheme is inherently prone to tampering and spying the activity/demands of other users. In this paper we develop schemes that are able to provide full information-theoretic privacy of the users' demands, while still providing a non-trivial caching gain. To motivate the reader and show that this is indeed possible, we continue our toy example with the following scheme, which is a special case of the MDS-based scheme in Theorem~\ref{thm:scheme 2}.  
In the placement phase, we encode each file $F_i$ by a $(4,3)$ MDS code (i.e., each file $F_i$ is split into $3$ blocks of $\Bsf/3$ bits each, which are then encoded by a $(4,3)$ MDS code such that each of the four  MDS coded  symbols has $\Bsf/3$ bits).   Each file can be reconstructed by  any three MDS coded symbols. The four MDS coded symbols are denoted by  $S^i_{1},S^i_{2},S^i_{3},S^i_{4}$. We  randomly   generate a permutation of $\{1,2,3,4\}$, denoted by $\pv_i=(p_{i,1},p_{i,2},p_{i,3},p_{i,4})$ and let  $F_{i,\emptyset}=S^i_{p_{i,1}}$, $F_{i,\{1\}}=S^i_{p_{i,2}}$, $F_{i,\{2\}}=S^i_{p_{i,3}}$, and $F_{i,\{1,2\}}=S^i_{p_{i,4}}$.
We let user $1$ cache   $F_{i,\{1\}}$  and $F_{i,\{1,2\}}$, and   user $2$ cache   $F_{i,\{2\}}$  and $F_{i,\{1,2\}}$. Notice that,  for the sake of successful decoding, user $1$ knows  the compositions of $F_{i,\{1\}}$ and $F_{i,\{1,2\}}$ (i.e., it knows from which code on which bits   $F_{i,\{1\}}$ and $F_{i,\{1,2\}}$  are generated),
but it does not know $ \pv_i$, i.e., it does not know which one of   $F_{i,\{1\}}$  and $F_{i,\{1,2\}}$  is cached by user $2$. Hence,  $F_{i,\{1\}}$  and $F_{i,\{1,2\}}$ are equivalent from the viewpoint of user $1$. Similarly, $F_{i,\{2\}}$ and $F_{i,\emptyset}$ are also  equivalent from the viewpoint of user $1$. 

In the delivery phase, we also consider two demands:
\begin{itemize}
\item if the demand is $(1,2)$, we transmit  $F_{1,\{2\}}\oplus F_{2,\{1\}}\oplus F_{3,\{1,2\}}$, such that user $1$ can recover $F_{1,\{2\}}$  and user $2$ can recover $F_{2,\{1\}}$. Notice that each user only knows 
the composition of each MDS coded symbol in the sum, without knowing whether the other user caches it or not.
 From the viewpoint of user $1$, in the sum  there is  one MDS coded symbol  from each file  and   among these  MDS coded symbols  it caches the ones from
the non-demanded files (i.e., $F_2$ and $F_3$).  
\item if the demand is $(1,1)$, we transmit  $F_{1,\emptyset}\oplus F_{2,\{1,2\}} \oplus F_{3,\{1,2\}}$, such that users $1$ and $2$ can recover $F_{1,\emptyset}$. Again, from the viewpoint of user $1$, in the sum  there is  one MDS coded symbol  from each file  and   among these  MDS  coded symbols, it caches the ones from the non-demanded files.
\end{itemize}
For the above two demands, 
from the viewpoint of user $1$, the delivery phases are equivalent. Hence, user $1$ cannot know any information about the request  of user $2$.
A symmetric situation holds for user $2$ and for all other possible demands.

In practice, 
 it may be important to preserve the privacy of  the users demands. The above example motivates the following question: what is the fundamental coded caching gain subject to such strict privacy constraint on the users demands? 
 In this paper, we focus on the private shared-link caching model with multiple requests from an information-theoretic viewpoint, where each user requests $\Lsf$ files. The objective is to design a private caching scheme with minimum load in the delivery phase, in order to
maintain the successful decoding for each user and also to prevent each user from getting any information about other users' demands.

 \subsection{Relation to Private Information Retrieval}  
\label{sub:PIR}
The privacy of the users demands was originally considered as the Private Information Retrieval (PIR) problem in~\cite{origPIR1995Chor}. In this setting,  a user wants to  retrieve a desired message from some distributed non-colluding databases (servers), and the objective is to prevent any server from  retrieving any information about the users' demand. Recently, the authors in~\cite{PIR2017Sun} characterized the information-theoretic capacity of the PIR problem by proposing a novel converse bound and a coded PIR scheme based on an interference alignment idea.

Later,   models combining the PIR problem with some caching component were proposed in~\cite{PSIPIR2017Chen,single2018Li,cachePIR2017Tandon,partial2017Wei,unknownPIR2017wei,storagePIR2018Attia}. 
In~\cite{PSIPIR2017Chen}, the user randomly caches some files in the library and its side information is unknown to the servers. The capacity region of the rate in terms of the number of cached files was characterized in~\cite{PSIPIR2017Chen}. The authors extended the model in~\cite{PSIPIR2017Chen} to the single-server multi-user case, where each user caches some files and  knows the demands of other users. A caching scheme based on 
Maximum Distance Separable (MDS) code was proposed.
In~\cite{cachePIR2017Tandon,partial2017Wei,unknownPIR2017wei}, for the single-user PIR problem with end-user-cache,  instead of caching the whole files, the user can choose any bits to store as in the coded caching model. 
Novel converse and achievable bounds were proposed in~\cite{cachePIR2017Tandon,partial2017Wei,unknownPIR2017wei} for the cases where the user's cache is known, partially known, and unknown to the servers, respectively. 
The authors in~\cite{storagePIR2018Attia} considered the   single-user PIR problem with end-database-caches, where each server can choose any bits to store instead of being able to access to the whole library. Under the constraint of uncoded cache placement, the optimal PIR scheme was given in~\cite{storagePIR2018Attia}.
The PIR problem was then generalized to the Private Computation (PC) problem in~\cite{sun2017pc}, where the  user should   compute a
function on the library instead of directly retrieving one message.

The considered coded caching problem with private demands aims to preserve the privacy of the demands of each user from other users, while the cache-aided PIR (or PC) problems aim to preserve the privacy of the users demands from the databases. Hence, the main challenge of the considered problem is to design multicast messages transmitted from the server such that each user can decode its desired files without getting any information of other users' demands,   while still achieving a non-trivial coded caching gain. 



\subsection{Contributions}  
\label{sub:contributions}
Our main contributions are as follows.
\begin{itemize}
 \item  {\bf Problem formulation.} We   formulate an information-theoretic shared-link coded caching model with multiple requests,  and the constraints on the information-theoretic privacy of the users demands from other users. 
 \item  {\bf Private coded caching schemes.} With   a novel idea of  private placement precoding  (which makes the cached (resp. uncached) bits from each file equivalent from the viewpoint of each user), we then  propose two private coded caching schemes with two different strategies  to generate a set of coded multicast messages which is symmetric over all the files (i.e., independent of the users' demands) from the viewpoint of each user.
\begin{enumerate}
\item Inspired by the virtual-user strategy originally introduced in~\cite{Engle2017privatecaching}, we propose a novel private caching scheme, referred to as {\it virtual user} scheme, by generating $\binom{\Nsf}{\Lsf} \Ksf-\Ksf$ virtual users such that each $\Lsf$-subset of files is demanded by exactly $\Ksf$ real or virtual (effective) users. We then propose a private delivery scheme based on   the $\binom{\Nsf}{\Lsf}\Ksf$-user MAN delivery scheme. Thus by `hiding' the real users among all effective users, the set of coded multicast messages   is symmetric over all the files and independent of the users' demands, from the viewpoint of each user. Notice that the caching scheme in~\cite{Engle2017privatecaching}  generates an arbitrary number of  virtual users each of whom randomly demands one file, which cannot  guarantee  the  information-theoretic   privacy constraint even if the number of virtual users goes to infinity. 
\item The main limitation of the virtual-user scheme is its sub-packetization level, which is equal to the sub-packetization level of the $\binom{\Nsf}{\Lsf}\Ksf$-user MAN coded caching scheme  (it has the order   $\Oc \left( 2^{\binom{\Nsf}{\Lsf}\Ksf }\right)$ when $\Msf \approx \Nsf/2$). In order to reduce the sub-packetization level, we propose the second scheme, referred to as {\it MDS-based} scheme. With a novel MDS-based cache placement, the main strategy is to       generate multicast messages in the delivery phase, such that  each multicast message contains one MDS-coded symbol from each file and thus is symmetric over all the $\Nsf$ files from the viewpoint of each user. There is no MDS-coded symbol appearing in two multicast messages, which makes   the set of all multicast messages also  symmetric over all the $\Nsf$ files.
 The needed sub-packetization level is $ \Oc\left(2^{\Ksf}\right)$, 
 which  reduces exponentially the one of the virtual-user scheme and is the same as the maximal sub-packetization level  of the $\Ksf$-user MAN coded caching scheme.
 \end{enumerate}
\item {\bf Order optimality results.} We summarize the order optimality results of the two proposed schemes in Table~\ref{table:opt}. In short, the virtual-user scheme is order optimal within a constant factor when $\Nsf \leq \Lsf\Ksf$, or when $\Nsf\geq \Lsf\Ksf$ and $\Msf \geq \Nsf/\Ksf$. In addition, when $\Msf \geq \Nsf/2$, the virtual-user scheme and the MDS-based scheme have the same order optimality results. 
 \end{itemize}

 
 \begin{table*}
\protect\caption{Order optimality factors of the virtual-user scheme and the MDS-based scheme.}\label{table:opt}
 \begin{center}
\begin{tabular}{|c|c|c|c|c|c|}
\hline 
\multicolumn{1}{|c|}{} & \multicolumn{2}{c|}{$\Nsf>\Lsf \Ksf$, $\Msf<\Nsf/2$} & \multicolumn{2}{c|}{$\Nsf \leq \Lsf\Ksf$,$\Msf<\Nsf/2$} & \multirow{2}{*}{$\Msf\geq \Nsf/2$}\tabularnewline
\cline{1-5} 
 & $\Lsf=1$ & $\Lsf>1$ & $\Lsf=1$ & $\Lsf>1$ & \tabularnewline
\hline 
Virtual-user scheme & $4$,  for $\frac{\Nsf}{\Ksf} \leq \Msf< \frac{\Nsf}{2} $ & $22$,  for $\frac{\Nsf}{\Ksf} \leq \Msf< \frac{\Nsf}{2}$ & $8$ & $22$ &   $2$ \tabularnewline
\hline 
MDS-based scheme &  &  &  &  &  $2$ \tabularnewline
\hline 
\end{tabular}
\par\end{center}
\end{table*}
\subsection{Paper Organization}
The rest of this paper is organized as follows.
Section~\ref{sec:model} formulates the considered shared-link  caching model with private demands.
Section~\ref{sec:main results} lists all the results in this paper and provide some numerical evaluations. 
Section~\ref{sec:general} presents the   proposed private caching schemes.  
Section~\ref{sec:conclusion} concludes the paper and some proofs are given in the Appendices.

\subsection{Notation Convention}
Calligraphic symbols denote sets, 
bold symbols denote vectors, 
and sans-serif symbols denote system parameters.
We use $|\cdot|$ to represent the cardinality of a set or the length of a vector;
$[a:b]:=\left\{ a,a+1,\ldots,b\right\}$ and $[n] := [1,2,\ldots,n]$; 
$\oplus$ represents bit-wise XOR; $\mathbb{E}[\cdot]$ represents the expectation value of a random variable; 
$[a]^+:=\max\{a,0\}$; 
we let $\binom{x}{y}=0$ if $x<0$ or $y<0$ or $x<y$;
we denote the power set of $[a]$ by $\text{Pow}(a)$, and sort all sets in       lexicographic   order. $\text{Pow}(a,j)$ denotes the $j^{\text{th}}$ set. For example, 
$$
\text{Pow}(3)=\{\emptyset,\{1\},\{1,2\}, \{1,2,3\}, \{1,3\},\{2\},\{2,3\},\{3\}\},
$$
and $\text{Pow}(3,1)=\emptyset$, $\text{Pow}(3,2)=\{1\}$, etc.


\section{System Model and Related  Results}
\label{sec:model}

\subsection{System Model}
\label{sub:system}
A  $(\Ksf,\Nsf,\Msf,\Lsf)$ shared-link caching system  with private demands is defined as follows. 
The system contains a server with access to a library of $\Nsf$ independent files, denoted by $(F_{1},F_{2},\dots,F_{\Nsf})$, where each file is composed of $\Bsf$ i.i.d. bits. As in~\cite{dvbt2fundamental}, we assume  that $\Bsf$ is sufficiently large such that any sub-packetization of the files is possible.
The server is connected to $\Ksf$ users through an error-free shared-link. The caching system operates in two phases.

{\it Placement Phase.}
 During the placement phase, user $k\in[\Ksf]$ stores content 
 in its cache of size $\Msf\Bsf$ bits without knowledge of later demands, where $\Msf\in[0,\Nsf]$. 
  We denote the content in the cache of user $k\in[\Ksf]$ by $Z_{k}$, which contains two parts 
\begin{align}
Z_{k}=(\mathscr{M}(C_k),C_k),
\end{align}  
 where  $C_k$ represents cached content from the $\Nsf$ files, 
\begin{align}
C_{k}=\phi_{k}(F_{1},\ldots,F_{N}, \mathscr{M}(C_k)), \label{eq:cK}
\end{align}  
 and  
   $\mathscr{M}(C_k)$ represents the metadata/composition of   $C_k$ (i.e., from which code on which bits,    $C_{k}$  are generated). For any bit in   $C_k$,   the metadata of this  bit does not reveal
which of the other users cache it. Notice that $\mathscr{M}(C_1),\ldots,\mathscr{M}(C_{\Ksf})$ are random variables over $\Cc_1,\ldots,\Cc_{\Ksf}$, representing all types of cache placement which can be used by the $\Ksf$ users.
In addition, for any $k\in [\Ksf]$, the realization of $\mathscr{M}(C_k)$ is known by user $k$   and is not   known by other users. 

We assume that the total length of $\mathscr{M}(C_k)$  compared to  the file length $\Bsf$ such that, for simplicity, the relevant cache size constraint is
\begin{align}
\frac{H(Z_{k})}{\Bsf}=\frac{H(C_k)}{\Bsf}\leq \Msf, \ \forall k \in [\Ksf].   \ \text{(Memory size)} \label{eq:memory size}
\end{align}  
  We also denote by $\mathbf{Z}:=(Z_{1},\ldots,Z_{K})$ the content of all $\Ksf$ caches.

 

{\it Delivery Phase.} 
 During the delivery phase, each user demands $\Lsf$ files, where $\Lsf\in [\Nsf]$. In this paper, we consider $\Nsf\geq \Lsf$ to ensure each user has $\Lsf$ demands.
 The demand vector of user  $k\in[\Ksf]$ are denoted by $\dv_k:=(d_{k,1},d_{k,2},\ldots,d_{k,\Lsf})$, where $1\leq d_{k,1}< d_{k,2}<\dots<d_{k,\Lsf}\leq \Nsf$. The demand matrix of all $\Ksf$ users is denoted by $\mathbb{D}:=[\dv_1;\dv_2;\ldots;\dv_{\Ksf}]$.
 In addition, we define  $\mathbb{D}_{\backslash\{k\}}$ for each $k \in [\Ksf]$ as the demand vectors of all users except user $k$, where  
\begin{align}
\mathbb{D}_{\backslash\{k\}}:= [\dv_1; \ldots; \dv_{k-1}, \dv_{k+1},\ldots, \dv_{\Ksf}].
\end{align} 
We also denote the set of all possible demand matrices by
\begin{align}
 \mathscr{D}:=\{\mathbb{D}:1\leq d_{k,1}< d_{k,2}<\dots<d_{k,\Lsf}\leq \Nsf, \forall k\in [\Ksf] \}.
\end{align}

We assume that   the metadatas of users' caches,    users' demands, and the library contents are independent, 
 \begin{align}
 &H\big(F_{1},F_{2},\dots,F_{\Nsf},\{\mathscr{M}(C_k):k\in [\Ksf]\},\{T_{\Sc}:\Sc\subseteq [\Ksf]\}, \mathbb{D} \big)\nonumber\\
 &=\Nsf \Bsf+ H(\{\mathscr{M}(C_k):k\in [\Ksf]\})+\sum_{\Sc\subseteq [\Ksf]} H(T_{\Sc})+H(\mathbb{D}).
 \end{align}

\begin{figure}
\centerline{\includegraphics[scale=0.6]{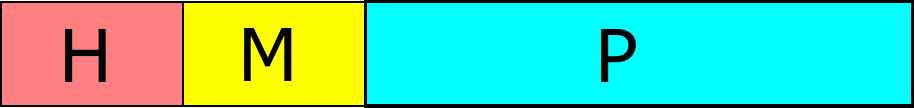}}
\caption{\small The delivery packet of $X_{\Sc}$, where `H' represents {\it Header}, `M' represents {\it Metadata}, `P' represents {\it Payload}.}
\label{fig: packet1}
\end{figure}

 Given the demand matrix $\mathbb{D}$ and the users' caches $\mathbf{Z}$,
the server broadcasts a packet $X=(\mathscr{M}(P), P)$ which includes three parts (Header,   Metadata, and Payload) as illustrated in Fig.~\ref{fig: packet1}. The header of  $X$ provides information (e.g., protocols, source, destination, etc.) to ensure that all users in $[\Ksf]$ can receive successfully the broadcasted packet  $X $.   
 To ensure the successful decoding on the payload, 
 the metadata $\mathscr{M}(P)$ represents the composition of the  payload $P$. 
 Notice that $\mathscr{M}(P)$ is random variable  over $\Pc$, representing all types of transmissions by the server.
The payload   contains the coded packets from the $\Nsf$ files, 
 \begin{align}
P=\psi\big(F_{1},\ldots,F_{\Nsf}, \mathscr{M}(P) \big).
\end{align}  
 We also assume that the total length of the header and  metadata are negligible compared to  the payload, such that we have 
\begin{align}
\Rsf:=H(X)/\Bsf=H(P)/\Bsf, 
\end{align}  
where $\Rsf$ represents the  load (i.e., normalized number of total transmitted bits) of $X$.
 
  The constraints  on the decoding of the demanded file by each user while maintaining the privacy is given as follows.
For each user $k\in [\Ksf]$, 
it must hold that
\begin{align}
H(\{F_{i}:i\in \dv_k\}|  X , Z_k, \dv_k  )=0, \ \forall k\in[\Ksf]. \ \text{(Decodability)}\label{eq:decodability}
\end{align}
 In addition,  given $\dv_k$, user $k$ cannot get any information about the demands of other users  from $X$,
i.e., the information-theoretic  privacy constraint is
\begin{align}
I(\mathbb{D}_{\backslash\{k\}} ;X, Z_k| \dv_k  )=0, \ \forall k\in[\Ksf].   \ \text{(Privacy)}\label{eq:privacy}
\end{align}
In other words, the mutual information between $\mathbb{D}_{\backslash\{k\}}$ and the user information after the delivery phase, quantifies in precise information-theoretic terms the information leakage of the delivery phase on the   demands of other users in the perspective of  user $k$. The privacy constraint in~\eqref{eq:privacy} (zero information leakage) corresponds to perfect secrecy in an information-theoretic sense (see~\cite[Chapter 22]{networkinformation}).

Since $Z_k$ is independent of $\mathbb{D}$,  the privacy constraint in~\eqref{eq:privacy} can be also written as 
\begin{align}
I(\mathbb{D}_{\backslash\{k\}} ; X| Z_k,\dv_k   )=0, \ \forall k\in[\Ksf].   \ \text{(Privacy)}\label{eq:privacy2}
\end{align}

{\it Objective.}
By the constraint of privacy, we can see that the transmitted loads for different demand matrices should be the same;  otherwise, the transmitted load which can be counted by each user will reveal information about the users demands.
The  memory-load tradeoff
 $(\Msf,\Rsf)$ is said to be achievable for the memory constraint $\Msf$, 
if there exist a two-phase private caching scheme as defined above  
such that all possible demand matrices can be delivered with load at most $\Rsf$ while the decodability and privacy constraints in~\eqref{eq:decodability} and~\eqref{eq:privacy2} are satisfied.
The objective is to determine, for a fixed $\Msf\in [0,\Nsf]$, the minimum load $\Rsf^{\star}$.

Notice that in the rest of the paper, when we introduce achievable schemes, we directly provide the construction of the payloads and skip the description on their metadatas. 

\subsection{MAN Coded Caching Scheme}
\label{sub:MAN}
In the following, we  recall the MAN shared-link caching scheme proposed in~\cite{dvbt2fundamental} and show this scheme cannot preserve the privacy of the users demands. We first focus on $\Lsf=1$, i.e., each user requests one file.

{\it Placement Phase.}
Let $\Msf=\Nsf t^{\prime}/\Ksf$ where $t^{\prime}\in [0:\Ksf]$.  Each file $F_i$ where $i\in[\Nsf]$ is divide into $\binom{\Ksf}{t^{\prime}}$ non-overlapping and equal-length subfiles, $F_{i}=\{F_{i,\Wc}:\Wc\subseteq [\Ksf], |\Wc|=t^{\prime}\}$, while each user $k\in [\Ksf]$ caches $F_{i,\Wc}$ where $k\in \Wc$. In other words, 
\begin{align}
C_k=\{F_{i,\Wc}: i\in [\Nsf], \Wc\subseteq [\Ksf], |\Wc|=t^{\prime}, k\in \Wc\}, \forall k\in [\Ksf]. \label{eq:MAN placement}
\end{align}
Hence, each user caches $\Nsf\Bsf\frac{\binom{\Ksf-1}{t^{\prime}-1}}{\binom{\Ksf}{t^{\prime}}}=\Msf \Bsf$ bits.

 {\it Delivery Phase.}
For each $\Sc\subseteq [\Ksf]$ where $|\Sc|=t^{\prime}+1$, the server generates an MAN multicast message
\begin{align}
X_{\Sc}=\underset{k\in \Sc}{\oplus } F_{ d_{k,1}, \Sc\setminus \{k\}}. \label{eq:MAN delivery}
\end{align} 
 The server transmits 
$X=\big(X_{\Sc}: \Sc\subseteq [\Ksf], |\Sc|=t^{\prime}+1)$.
 {\bf In this paper,  we define the composition of an XOR message of subfiles (or MDS coded symbols) as the containing subfiles (or MDS coded symbols)  in this message.} 
It can be seen that in the composition of $X_{\Sc}$, each user in $\Sc$ caches all subfiles except $ F_{ d_{k,1}, \Sc\setminus \{k\}}$ such that it can recover $ F_{ d_{k,1}, \Sc\setminus \{k\}}$. Considering all  $\Sc\subseteq [\Ksf]$ where $|\Sc|=t^{\prime}+1$, each user can recover its desired file, i.e., the decodability constraint in~\eqref{eq:decodability} is satisfied. 

When $\Lsf>1$, the transmission is divided into $\Lsf$ rounds, where in each round we serve one demand of each user. By using the above MAN delivery scheme by $\Lsf$ times,
the achieved load by the $\Lsf$-round MAN coded caching scheme   is as follows,
 \begin{align}
 (\Msf_{\text{MAN}},\Rsf_{\text{MAN}})=\left(\frac{\Nsf t^{\prime}}{\Ksf},  \Lsf \frac{\Ksf-t^{\prime}}{t^{\prime}+1}  \right), \  \forall t^{\prime}\in [0:\Nsf]. \label{eq: MAN load} 
 \end{align}
For other memory sizes, we can take the lower convex envelope of the corner points in~\eqref{eq: MAN load}.

However, consider one $\Sc\subseteq [\Ksf]$ where $|\Sc|=t^{\prime}+1$, each user knows the metadata of each subfile in $X_{\Sc}$. Hence, it knows the composition of $X_{\Sc}$, i.e, it knows the union set of the demanded files by users in $\Sc$ is $\cup_{k\in \Sc} \{ d_{k,1}\}$, which contradicts the privacy constraint in~\eqref{eq:privacy2}. 
Even if we   hide the identity of the intended users of each multicast messages (i.e., each user  $k\in \Sc$ does not know that $X_{\Sc}$ is useful to users in $\Sc\setminus \{k\}$), 
the composition of the set of all multicast messages is not symmetric over all the $\Nsf$ files if the number of users demanding each file is not the same.





\section{Main Results}
\label{sec:main results}
In this section, we   list the proposed results of this paper for the considered problem described in Section~\ref{sub:system}, and then provide some numerical evaluations.

We first provide a 
baseline scheme, which trivially  uses uncoded caching to let   each user   recover the whole library.  
\begin{thm}[Baseline Scheme]
\label{thm:scheme 1}
For the $(\Ksf,\Nsf,\Msf)$ shared-link caching system with private demands, $\Rsf^{\star}$ is upper bounded by  
\begin{align}
\Rsf^{\star}\leq \Rsf_{\text{base}}=\Nsf-\Msf. 
\label{eq:scheme 1}
\end{align}
\hfill $\square$ 
\end{thm}
\begin{IEEEproof} 
 {\it Placement Phase.}
Each user caches the same $\Msf\Bsf/\Nsf$ bits of each file $F_i$ where $i\in [\Nsf]$. We denote the cached part of $F_i$ by $F^{\text{c}}_{i}$ and the uncached part by $F^{\text{u}}_{i}$. 
Since users have the same cached content, each user knows the cached content of other users.

{\it Delivery Phase.}
 The server transmits $X=\{F^{\text{u}}_{i}:i\in [\Nsf]\}$.  For the decodability, each user  has received uncached part of each  file in the library, which includes its desired files.  For the privacy, 
since we transmit the   the uncached parts of all files, each user cannot know which files among them are desired by other users. 
  Hence, the privacy of the users demands   is preserved.

{\it Performance.}
The normalized length of the uncached part of each file is $1-\frac{\Msf}{\Nsf}$. Hence, the achieved load is $ \Nsf \left(1-\frac{\Msf}{\Nsf} \right)=\Nsf-\Msf$ as in Theorem~\ref{thm:scheme 1}.
\end{IEEEproof}

  In order to use the coded caching strategy while preserving the privacy of the users' demands, we aim to design private caching schemes such that the composition of the set of all   multicast messages is symmetric over all the $\Nsf$ files from the viewpoint of each user. For this purpose,  with a novel idea of  private placement precoding summarized in Remark~\ref{rem:generation to any placement} (which makes the cached (resp. uncached) bits  from each file equivalent from the viewpoint of each user), we propose  two private caching schemes,  the {\it virtual-user } scheme and the {\it MDS-based scheme} scheme, based on  two different   strategies, respectively.  The main ingredients of
the two schemes are as follows. 
\begin{enumerate}
\item Virtual-user scheme.  Since in the library there are $\Nsf$ files while each user demands $\Lsf$ among them (i.e., the demand set of each user contains $\Lsf$ files), it can be seen that there are totally $\binom{\Nsf}{\Lsf}$ possibilities of demand sets, each of which is requested by at most $\Ksf$ users. Hence,   we can generate $\binom{\Nsf}{\Lsf} \Ksf-\Ksf$ virtual users such that the system contains totally $\binom{\Nsf}{\Lsf}\Ksf$  effective users (i.e., real or virtual users) and each possible demand set is requested by exactly $\Ksf$ effective users.  
We then use the MAN delivery scheme over these $\binom{\Nsf}{\Lsf}\Ksf$  effective users. 
{\bf To conclude, the strategy is that even if the composition of each multicast message is not symmetric over all the $\Nsf$ files, with the fact that  the number of effective users demanding  each  file is identical, we let the composition of the set of all multicast messages be symmetric over all the $\Nsf$ files from the viewpoint of each user.}
\item  MDS-based scheme. {\bf Different from the first strategy, the second strategy is letting each multicast message be symmetric over all the $\Nsf$ files from the viewpoint of each user.} More precisely,
with a novel  private MDS-coded cache placement,  
we generate symmetric multicast messages in the delivery phase, such that   each multicast message (assumed to be useful to users in $\Sc$) contains one MDS-coded symbol from each file, where each user $k\in \Sc$ caches all MDS-coded symbols from the files which it does not require.   As a result, the composition of the   multicast   messages is equivalent for different demands from the viewpoint of each user.   With some careful design, there is no MDS-coded symbol appearing in two multicast messages, which makes  the composition of the set of all multicast messages also  symmetric over all the $\Nsf$ files. 
\end{enumerate}
 

 The achieved load of the virtual user scheme is given in the following, whose proof  could be found in Section~\ref{sub:extended scheme}.
\begin{thm}[Virtual-user  scheme]
\label{thm:extended scheme}
For the $(\Ksf,\Nsf,\Msf,\Lsf)$ shared-link caching system with private demands, $\Rsf^{\star}$ is upper bounded by $\Rsf_{\text{v}}$, where the memory-load tradeoff $(\Msf,\Rsf_{\text{v}})$ is  the lower convex envelope
of $(0,\Nsf)$ and the following memory-load pairs
\begin{align}
&  \left(\frac{t}{\binom{\Nsf}{\Lsf}\Ksf}\Nsf,
\Lsf\frac{\binom{\Nsf}{\Lsf}\Ksf-t}{t+1} \right), \ \forall  t\in \left[ \binom{\Nsf}{\Lsf}\Ksf \right].
\label{eq:extended scheme} 
\end{align}
\hfill $\square$ 
\end{thm}
 Notice that the idea to introduce virtual user to  hide  the demands of the real users was originally proposed in~\cite{Engle2017privatecaching}.~\cite{Engle2017privatecaching}  focuses on the case of single request (i.e., $\Lsf=1$) and generates an arbitrary number of  virtual users each of whom randomly demands one file, which cannot  guarantee  the  information-theoretic   privacy constraint in~\eqref{eq:privacy} even if the number of virtual users goes to infinity. Instead, we propose  rigorous  code constructions  on novel private placement and delivery phases, such that the   information-theoretic privacy constraint in~\eqref{eq:privacy} holds.  In the proposed virtual-user scheme we introduce  a finite and fixed number of virtual users which depends on the system parameters, and   a determinate scenario to choose one demand set for each virtual user.

Compared to   the existing converse bound  in~\cite{improvedlower2017Ghasemi,yas2,Sengupta2017multirequest} for the  shared-link caching model without privacy, we have the following order optimality results of the virtual-user scheme which will be proved in Appendix~\ref{sub:proof of general order}. 
\begin{thm}[Order Optimality]
\label{thm:extended order}
For the $(\Ksf,\Nsf,\Msf,\Lsf)$ shared-link caching system with private demands,
\begin{itemize}
\item if $\Lsf=1$, the virtual-user scheme in  Theorem~\ref{thm:extended scheme}  is order  optimal within a factor of $8$ when $\Nsf \leq \Ksf$, and of $4$ when $\Nsf >\Ksf $ and $\Msf \geq \Nsf/\Ksf$;
\item if $\Lsf>1$, the virtual-user scheme in  Theorem~\ref{thm:extended scheme}  is order  optimal within a factor of  $22$ when $\Nsf \leq \Lsf\Ksf$, or when $\Nsf > \Lsf\Ksf $ and $\Msf \geq \Nsf/\Ksf$.
\end{itemize} 
\hfill $\square$ 
\end{thm}
 Intuitively, the order optimality results arise from the fact that introducing virtual users does not increase much load when the memory size is not small (a similar observation was originally pointed out in~\cite{Engle2017privatecaching}).

The virtual-user scheme in  Theorem~\ref{thm:extended scheme} contains $\binom{\Nsf}{\Lsf}\Ksf$ effective users, and generate a subfile of each file which is then cached by effective users in $\Sc$ for each $\Sc \subseteq [\Ksf]$ where $|\Sc|=t$. Hence, the    needed   sub-packetization level is  
$$
\binom{\binom{\Nsf}{\Lsf}\Ksf}{t} \approx 2^{\binom{\Nsf}{\Lsf}\Ksf \Hc(\Msf/\Nsf)}
$$
where $\Hc(p)=-p \log_2(p)-(1-p)\log_2(1-p) $ is the binary entropy function. Hence,   the maximal  sub-packetization level of the virtual-user scheme (when $\Msf/\Nsf \approx 1/2$) is  
exponential to $\binom{\Nsf}{\Lsf}\Ksf$ (i.e.,  $\Oc\left(2^{ \binom{\Nsf}{\Lsf} \Ksf} \right)$), which is much higher than the maximal sub-packetization level  of the $\Ksf$-user MAN coded caching scheme without virtual users (exponential to $\Ksf$, i.e.,  $\Oc\left(2^{ \Ksf}\right)$).
 To enable the application of the private caching scheme in the practice, it is important to  reduce the sub-packetization level (at least the maximal sub-packetization level should  not  be exponentially larger than the original MAN coded caching scheme). Hence, we propose the MDS-based scheme
  with sub-packetization level   $  \Oc\left(2^{ \Ksf}\right)$. 
  The detailed description of the MDS-based scheme and the proof of its achieved load can be found in Sections~\ref{sub:improved 1} and~\ref{sub:improved 2}.
\begin{thm}[MDS-based scheme]
\label{thm:scheme 2}
For the $(\Ksf,\Nsf,\Msf,\Lsf)$ shared-link caching system with private demands, $\Rsf^{\star}$ is upper bounded by    $\Rsf_{\text{m}}$, where  the memory-load tradeoff $(\Msf, \Rsf_{\text{m}})$ is 
the lower convex envelope
of $(0,\Nsf)$, and the following memory-load pairs
\begin{align}
&  \left(\Nsf \frac{ 2^{\Ksf-1} }{2^{\Ksf-1}+\binom{\Ksf-1}{t}+\binom{\Ksf-1}{t+1}+\dots+\binom{\Ksf-1}{\Ksf-1}} ,
\Lsf\frac{2^{\Ksf}-\binom{\Ksf}{0}-\binom{\Ksf}{1}-\dots-\binom{\Ksf}{t} }{2^{\Ksf-1}+\binom{\Ksf-1}{t}+\binom{\Ksf-1}{t+1}+\dots+\binom{\Ksf-1}{\Ksf-1}} \right) , \ \forall  t\in [0:\Ksf],
\label{eq:scheme 2 1}\\
&\text{and } \left(\frac{2\Ksf-1}{2\Ksf}\Nsf, \frac{\Lsf}{2\Ksf} \right) .\label{eq:scheme 2 2}
\end{align}
\hfill $\square$ 
\end{thm}

Compared to   the existing converse bound  in~\cite{improvedlower2017Ghasemi,yas2,Sengupta2017multirequest} for the  shared-link caching model without privacy, 
the virtual-user scheme and the MDS-based scheme have the same order optimality results   when   $\Msf \geq \Nsf/2$, which will be proved in Appendix~\ref{sub:thm:general order}.
 \begin{thm}[Order Optimality]
\label{thm:general order}
For the $(\Ksf,\Nsf,\Msf,\Lsf)$ shared-link caching system with private demands, when  $\Msf\geq \Nsf/2$,  both of $\Rsf_{\text{v}}$  and $\Rsf_{\text{m}}$ are order  optimal within a factor of $2$.
\hfill $\square$ 
\end{thm}
 

By  
comparing  the achievable bounds in~\eqref{eq:scheme 2 1}  (letting $t=\Ksf-1$) and~\eqref{eq:scheme 2 2}, with
 the converse bound for     the  MAN shared-link caching model with multiple requests in~\cite[Theorem 1]{franarxiv} (letting  $s=1$), 
 we have the following exact optimality result. 
\begin{thm}[Exact Optimality]
\label{thm:general opt}
For the $(\Ksf,\Nsf,\Msf,\Lsf)$ shared-link caching system with private demands where $\Msf \geq \min\left\{\frac{2\Ksf-1}{2\Ksf},\frac{2^{\Ksf-1}}{2^{\Ksf-1}+1}  \right\}\Nsf $, we have 
\begin{align}
\Rsf^{\star}=\Rsf_{\text{m}}=\Lsf\left(1-\frac{\Msf}{\Nsf} \right).\label{eq:exact optimal}
\end{align} 
\hfill $\square$ 
\end{thm}


 It can be seen that for the considered large memory size regime in Theorem~\ref{thm:general opt}, our proposed schemes can maintain the exact optimality for the   shared-link caching model with multiple requests, while preserving the privacy of the users demands.
Notice that for this purpose, 
the virtual-user scheme needs the memory size no less than $\frac{\binom{\Nsf}{\Lsf}\Ksf-1}{\binom{\Nsf}{\Lsf}\Ksf}\Nsf$.

From   Theorems~\ref{thm:extended order} and~\ref{thm:general order}, 
the only open case, where the multiplicative gaps between the proposed schemes and the existing converse bounds for the shared-link caching model without privacy constraint are not constant, is when  $\Nsf<\Lsf\Ksf$ and $\Msf< \Nsf/\Ksf.$

\begin{figure}
    \centering
    \begin{subfigure}[t]{0.5\textwidth}
        \centering
        \includegraphics[scale=0.6]{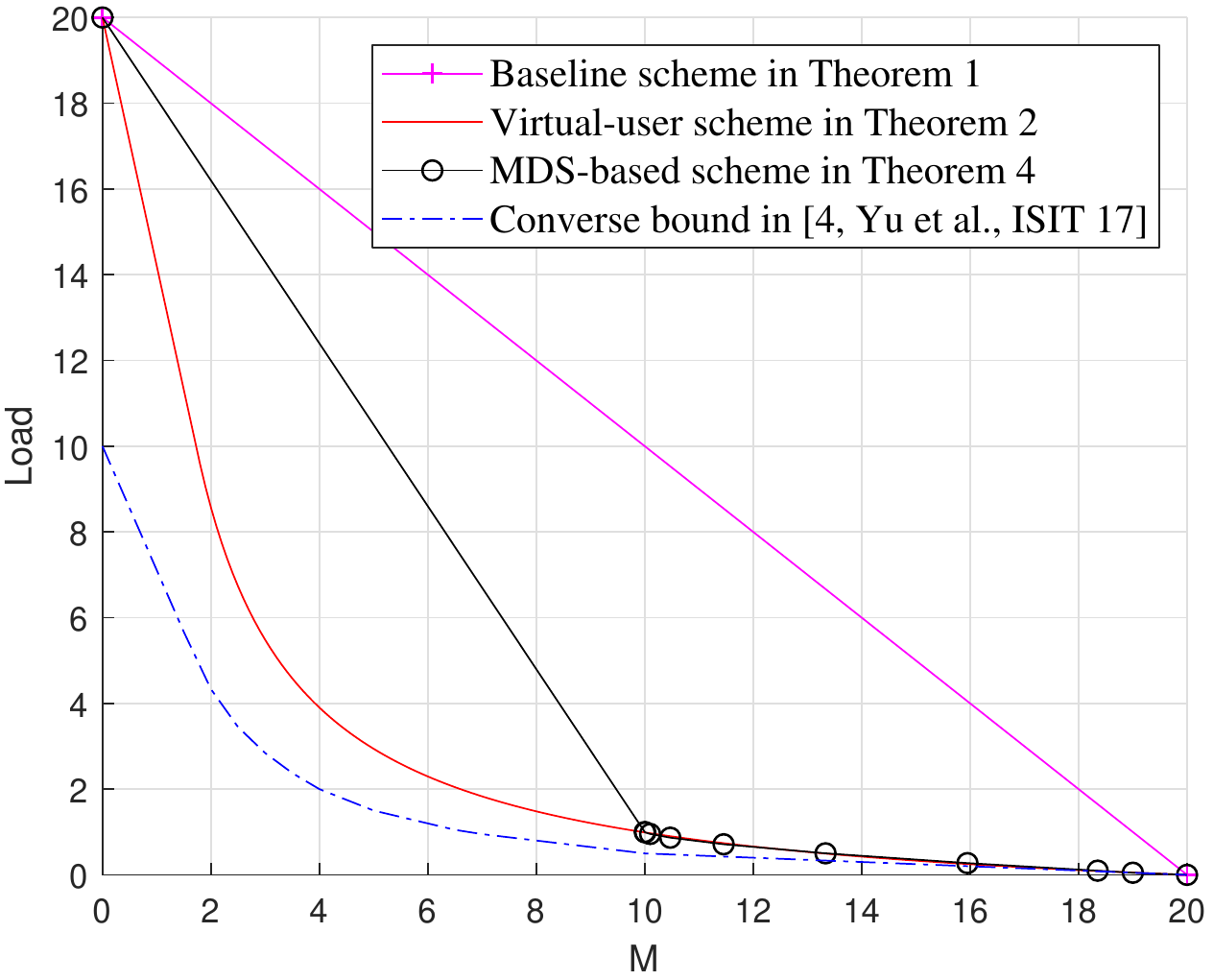}
        \caption{\small $(\Ksf,\Nsf,\Lsf )=(10,20,1)$.}
        \label{fig:extend num 1a}
    \end{subfigure}%
    ~ 
    \begin{subfigure}[t]{0.5\textwidth}
        \centering
        \includegraphics[scale=0.6]{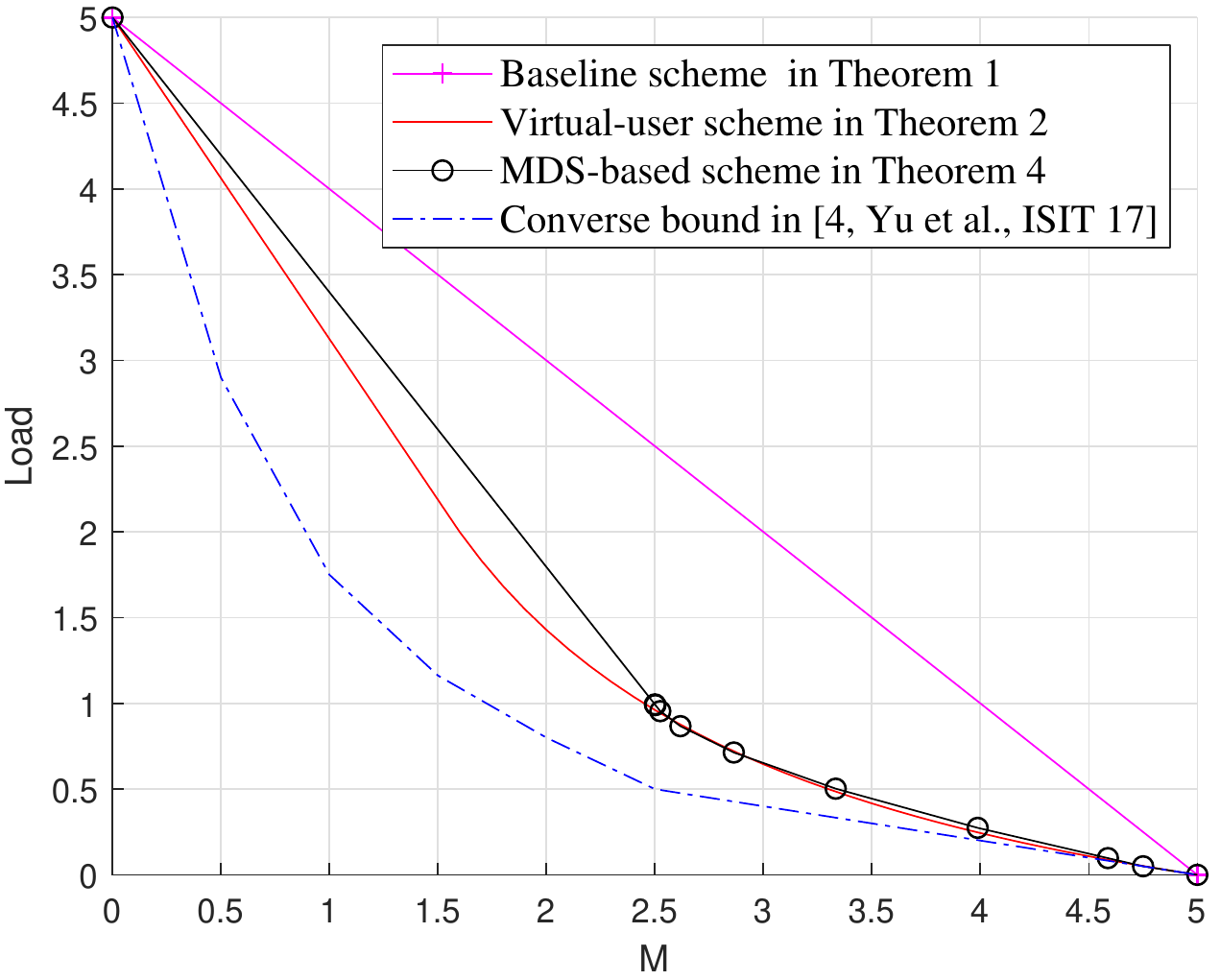}
        \caption{\small $(\Ksf,\Nsf,\Lsf)=(10,5,1)$. }
        \label{fig:extend num 1b}
    \end{subfigure}
    \caption{\small $(\Msf, \Rsf)$ tradeoff for the $(\Ksf,\Nsf,\Msf,\Lsf)$ shared-link caching system with private  demands.}
    \label{fig:extend num}
\end{figure}

 Finally, 
we provide numerical evaluations of the proposed private caching schemes for the $(\Ksf,\Nsf,\Msf,\Lsf)$ shared-link caching system with private   demands. In Fig.~\ref{fig:extend num} we let   $\Lsf=1$ and use the converse bound in~\cite{yas2} for the shared-link caching model with single request, as the converse bound in our problem.
In Fig.~\ref{fig:extend num 1a}, we let $(\Ksf,\Nsf)=(10,20)$ and in Fig.~\ref{fig:extend num 1b} we let $(\Ksf,\Nsf)=(10,5)$. Both of the figures show  that 
the   virtual-user scheme and the MDS-based scheme  outperform   the baseline scheme.  When $\Msf < \Nsf/2$, it can be seen that the  achieved load by the virtual-user scheme is lower than the MDS-based scheme. 
In addition,   when $\Msf \geq \Nsf/2$,
the    achieved loads by the  virtual-user scheme and the MDS-based scheme are close; in this regime, none of them always has the lower load than the other.

\section{Coded Caching with Private  Demands}
\label{sec:general}

 \subsection{Proof of Theorem~\ref{thm:extended scheme}} 
 \label{sub:extended scheme}  
 In the following, we describe the virtual-user scheme which achieves the memory-load tradeoff in~\eqref{eq:extended scheme}.
  We focus on one $t\in \left[ \binom{\Nsf}{\Lsf}\Ksf \right]$.
  We define that $\Usf:=\binom{\Nsf}{\Lsf}\Ksf$.

{\it Placement Phase.}
Each file $F_i$ where $i\in [\Nsf]$ is divided into $\binom{\Usf}{t} $ non-overlapping and equal-length pieces, denoted by $S^i_{1},\ldots,S^i_{\binom{\Usf}{t}}$, where each piece has $\frac{\Bsf}{\binom{\Usf}{t}}$ bits.  We  randomly  generate a permutation of $\binom{\Usf}{t}$, denoted by $\pv_i=(p_{i,1},\ldots,p_{i,\binom{\Usf}{t}})$,   independently and uniformly over the set of all possible permutations.
We sort all sets $\Wc\subseteq [\Usf]$ where $|\Wc|=t$,  in a lexicographic order, denoted by $\Wc(1),\dots, \Wc\left(\binom{\Usf}{t} \right)$.
For each $j\in \left[\binom{\Usf}{t} \right]$, we   generate a subfile  
\begin{align}
f_{i, \Wc(j)}= S^i_{p_{i,j}}.\label{eq:extend assignement 1}
\end{align}
Each user $k\in [\Ksf]$ caches $f_{i,\Wc}$ where $\Wc\subseteq [\Usf]$,  $|\Wc|=t$, and $k\in \Wc$.  
 Hence, each user   caches  $\binom{\Usf-1}{t-1} $ subfiles of each file,  and thus it totally caches $\frac{\binom{\Usf-1}{t-1} }{\binom{\Usf}{t} }\Nsf\Bsf  =\frac{t}{\Usf}\Nsf\Bsf=\Msf\Bsf$ bits, satisfying the memory size constraint in~\eqref{eq:extended scheme}. 

For each subfile of $F_i$   cached by  user $k\in [\Ksf]$, since the random permutation $\pv_i$ is unknown to user $k$, it does not know the other users who also cache it. Hence, each cached subfile of $F_i$ is equivalent from the viewpoint of user $k$. Similarly, each uncached subfile of $F_i$ is also equivalent from the viewpoint of user $k$.
 Hence,  from the viewpoint of user $k\in [\Ksf]$, each cached subfile of $F_i$ is equivalent from the viewpoint of user $k$, while each uncached subfile of $F_i$ is also equivalent.

 {\it Delivery Phase for $\mathbb{D}$.}
Recall that for one possible demand  vector by one user  $\dv:=(d_{1},d_{2},\ldots,d_{\Lsf})$, we should have  $1\leq d_{1}< d_{2}<\dots<d_{\Lsf}\leq \Nsf$.
Hence, there are totally $\binom{\Nsf}{\Lsf}$ possible demand vectors, denoted by $\dv^1,\ldots,\dv^{\binom{\Nsf}{\Lsf}}$. 
 We define that 
\begin{align}
n_j:=|\{k\in[\Ksf]:\dv_k=\dv^j \}|
\end{align} 
 where $j\in \left[ \binom{\Nsf}{\Lsf}\right]$, representing the number of real users demanding the demand vector $\dv^j$.
We then allocate one demand vector to each of the $\Usf-\Ksf$ virtual users as follows. 
For each $j\in \left[ \binom{\Nsf}{\Lsf}\right]$, we let $\dv_{1+j\Ksf-\sum_{q\in [j-1]}n_q}=\dots=\dv_{(j+1)\Ksf-\sum_{q\in[j] }n_q}=\dv^j$.
For example, when $j=1$, we let  $\dv_{\Ksf+1}=\dots=\dv_{2\Ksf-n_1}=\dv^1$;
  when $j=2$, we let $\dv_{2\Ksf-n_1+1}=\dots=\dv_{3\Ksf-n_1-n_2}=\dv^2$. 
  Hence, by this way, each possible demand  vector is requested by $\Ksf$ effective (real or virtual) users.

Recall for any $k\in [\Usf]$, we define $\dv_k=(d_{k,1},\ldots,d_{k,\Lsf})$.  In addition,  we define $\mathbb{G}_{a \times b}$ as the $a \times b$  parity-check matrix of the $[b ,b -a,a+1]$ MDS code (or an $a \times n$ Cauchy matrix) 
 such that each $ a$ columns are linearly independent (see~\cite{detailledisit}).
For each  set $\Sc \subseteq [\Usf]$ where $|\Sc|= t+1$, we generate 
\begin{align}
X_{\Sc}= \mathbb{G}_{\Lsf \times \Lsf(t+1)} \ [\fv_{k_1,\Sc \setminus\{k_1\}}; \fv_{k_2,\Sc \setminus\{k_2\}}; \dots; \fv_{k_{t+1} ,\Sc \setminus\{k_{t+1}\}} ].\label{eq:extend}
\end{align}
containing $\Lsf$ combinations 
where $(k_1,\ldots,k_{t+1})$ is a random permutation of $\Sc$  independently and uniformly over the set of all possible permutations,
and we define 
\begin{align}
\fv_{k_j,\Sc \setminus\{k_j\}}:=[f_{d_{k_j,1},\Sc \setminus\{k_j\}};f_{d_{k_j,2},\Sc \setminus\{k_j\}};\dots;f_{d_{k_j,\Lsf},\Sc \setminus\{k_j\}}], \ \forall j\in[t+1]. \label{eq:def of fv}
\end{align}
 
Finally, we  randomly  generate a permutation of $\left[\binom{\Usf}{t+1} \right]$, denoted by $\qv=(q_{1},\ldots,q_{\binom{\Usf}{t+1}})$, independently and uniformly over the set of all possible permutations.
We sort all sets $\Sc \subseteq [\Usf] $ where  $|\Sc |= t+1 $, in a lexicographic order, denoted by $\Sc(1), \ldots, \Sc\left(\binom{\Usf}{t+1} \right)$.
The server transmit 
\begin{align}
X=\left(X_{\Sc(q_{1})},\ldots,X_{\Sc \left(q_{\binom{\Usf}{t+1}}\right)} \right).\label{eq:extend X_K}
\end{align}

{\it Decodability.}
We focus on user $k\in [\Ksf]$.
From the metadata in $X$ (i.e., $\mathscr{M}(P)$), for each $j\in \left[\binom{\Usf}{t+1} \right]$, user $k\in [\Ksf]$ checks $X_{\Sc(q_{j})}$. If  $X_{\Sc(q_{j})}$ contains $\Lsf t$ cached subfiles  from the files not requested by user $k$, and $\Lsf$ subfiles from the files requested by user $k$, user $k$ knows $X_{\Sc(q_{j})}$ is useful to it and then decodes  the $\Lsf$ requested subfiles from the $\Lsf$ linear combinations in  $X_{\Sc(q_{j})}$, because  any $\Lsf$ columns in $\mathbb{G}_{\Lsf \times \Lsf(t+1)} $ are linearly independent.

After considering all transmitted packets in $X$, user $k\in [\Ksf]$ can recover all requested subfiles to reconstruct its requested files.

{\it Privacy.}
By the symmetric construction, from the viewpoint of each user  $k\in [\Ksf]$, for any demand matrix where user $k$ demands $\dv_k$, there are always $\Ksf$ effective users demanding each possible demand vector. In addition, since the placement permutations (i.e., $\pv_i$ where $i\in [\Nsf]$) is unknown to user $k$, the cached content of   each of the other  $\Usf-1$ effective users is equivalent from the viewpoint of user $k$. Hence, the composition of $X $ is totally equivalent for different demand matrices from the viewpoint of each user $k\in [\Ksf]$. 
 In other words, given $\dv_k$ and $Z_k$, it can be seen that $X$ is independent of $\mathbb{D}.$ Thus the proposed scheme is information-theoretically private.

 {\it Performance.}
For any demand matrix, we transmit $\binom{\Usf}{t+1} $   messages, each of which contains  $\frac{\Lsf \Bsf}{\binom{\Usf}{t} }$ bits. Hence, the achieved  load   is $\Lsf \frac{\binom{\Usf}{t+1} }{\binom{\Usf}{t}}=\Lsf\frac{\Usf-t}{t+1}$, as shown in~\eqref{eq:extended scheme}.
 The sub-packetization level is $\binom{\Usf}{t}$.

\subsection{Proof of~\eqref{eq:scheme 2 1}}
\label{sub:improved 1}
In the following we introduce the MDS-based scheme to achieve~\eqref{eq:scheme 2 1}. We first use a more complicated example than the toy example in Section~\ref{sub:private demand} to highlight more insights.

\begin{example}[$\Ksf=3$, $\Nsf=6$, $\Msf=3$, $\Lsf=2$]
\label{exp:improved scheme}
\rm
Consider a $(\Ksf,\Nsf,\Msf,\Lsf)=(3,6,3,2)$ shared-link caching problem with private demands. 

{\it Placement Phase.}
From~\eqref{eq:scheme 2 1}, we  can compute $t=0$ in this example. 
Each file $F_i$ where $i\in [\Nsf]$ is divided into $\binom{\Ksf}{0}+\dots+\binom{\Ksf}{\Ksf}=2^{\Ksf}=8$ non-overlapping and equal-length pieces, denoted by $S^i_{1},\ldots,S^i_{2^{\Ksf}}$, each of which contains $\frac{\Bsf}{8}$ bits. In this example where $t=0$, we do not need  the MDS precoding in the placement, which is necessary for $t>1$ and will be clarified in the next example. 
We  randomly  generate a permutation of $\left[2^{\Ksf} \right]$, denoted by $\pv_i=(p_{i,1},\ldots,p_{i,2^{\Ksf}})$, independently and uniformly over the set of all possible permutations.
We then assign each piece to a subfile according to  $\pv_i$ as follows,
\begin{align}
&f_{i,\emptyset}= S^i_{p_{i,1}}, \ f_{i,\{1\}}= S^i_{p_{i,2}}, \ f_{i,\{1,2\}}= S^i_{p_{i,3}}, \ f_{i,\{1,2,3\}}= S^i_{p_{i,4}} , \nonumber\\
& f_{i,\{1,3\}}= S^i_{p_{i,5}} ,\ f_{i,\{2\}}= S^i_{p_{i,6}}, \ f_{i,\{2,3\}}= S^i_{p_{i,7}}, \ f_{i,\{3\}}= S^i_{p_{i,8}}. \label{eq:example assignment}
\end{align}
Each user $k\in [\Ksf]$ caches $f_{i,\Wc}$ if $k\in \Wc$, i.e., 
the cached contents of the three users for each file $F_i$ are as follows:
\begin{itemize}
\item  User $1$ stores $f_{i,\{1\}}$, $f_{i,\{1,2\}}$, $f_{i,\{1,3\}}$,  and $f_{i,\{1,2,3\}}$. 
\item  User $2$ stores $f_{i,\{2\}}$, $f_{i,\{1,2\}}$, $f_{i,\{2,3\}}$,  and $f_{i,\{1,2,3\}}$. 
\item   User $3$ stores $f_{i,\{3\}}$, $f_{i,\{1,3\}}$, $f_{i,\{2,3\}}$,  and $f_{i,\{1,2,3\}}$. 
\end{itemize} 
 In addition, for each subfile of $F_i$ cached by  user $k\in [\Ksf]$, since the random permutation $\pv_i$ is unknown to user $k$, it does not know the other users who also cache it. Hence, each cached subfile of $F_i$ is equivalent from the viewpoint of user $k$. Similarly, each uncached subfile of $F_i$ is also equivalent from the viewpoint of user $k$.

Since each user caches $4$ subfiles (each of which has $\Bsf/8$ bits) for each file in its cache, it totally caches
 $
\Nsf\frac{4 \Bsf}{8} =3\Bsf=\Msf\Bsf
$
bits satisfying the memory size constraint.

 For the delivery phase, we do not consider all possible non-equivalent demand configurations, for the sake of brevity. Instead, we give two explicit examples of  the construction of the delivery phase and then extract some general properties that demonstrate the privacy.

{\it Delivery Phase for $\mathbb{D}=[1,2;3,4;5,6]$.}
For this demand matrix, user $1$ demands $F_{1}$ and $F_{2}$, user $2$ demands $F_3$ and $F_4$, and user $3$ demands $F_5$ and $F_6$. For each file $F_i$ where $i\in [\Nsf]$, we define 
\begin{align}
\Qc_{i}:= \{k\in [\Ksf]: i\in \dv_k\}, \label{eq:def of Q}
\end{align}
as the set of users demanding $F_i$. For $\mathbb{D}=[1,2;3,4;5,6]$, we have $\Qc_{1}=\Qc_2=\{1\}$, $\Qc_3=\Qc_4=\{2\}$, and $\Qc_5=\Qc_6=\{3\}$.

For each  subset $\Sc \subseteq [\Ksf]$ where $|\Sc|\geq t+1 =1$, we generate a multicast message $X_{\Sc}$ which is useful to the users $\Sc$. 
   Our purpose is to let $ X_{\Sc}$ be $\Lsf=2$ linear combinations of $\Nsf$ subfiles, where each file has one subfile in  $ X_{\Sc}$ and each user in $\Sc$ caches $\Nsf-\Lsf$ subfiles from the files which it does not request. In addition,  the $\Lsf=2$ linear combinations are generated by  $\mathbb{G}_{\Lsf \times \Nsf}$ where each $ \Lsf$ columns are linearly independent, such that each user  in $\Sc$ can recover the $\Lsf$ uncached subfiles.
 For each file $F_i$ where $i\in [\Nsf]$, the subfile of $F_i$ in $X_{\Sc}$ is $f_{i, \Sc\cup\Qc_i \setminus (\Sc \cap \Qc_i)}$ (the motivation of this construction will be explained in Remark~\ref{rem:non overlap}), which is cached by each user in $\Sc$ not requesting $F_i$, and not cached  by each user in $\Sc$ requesting $F_i$. 
   

We first consider $\Sc=\{1\}$, which only contains one user. We have 
\begin{align}
X_{\{1\}}=  \mathbb{G}_{2 \times 6} \ [f_{1,\emptyset};f_{2,\emptyset};f_{3,\{1,2\}};f_{4,\{1,2\}};f_{5,\{1,3\}};f_{6,\{1,3\}} ]  .\label{eq:P1}
\end{align}
From $X_{\{1\}}$, user $1$ caches all except $f_{1,\emptyset}$ and $f_{2,\emptyset}$, such that it can recover those two subfiles in $X_{\{1\}}$ (recall each two columns of $\mathbb{G}_{2 \times 6}$ are linearly independent).
Similarly, we have 
\begin{align}
&X_{\{2\}}=  \mathbb{G}_{2 \times 6} \ [f_{1,\{1,2\}};f_{2,\{1,2\}};f_{3,\emptyset};f_{4,\emptyset};f_{5,\{2,3\}};f_{6,\{2,3\}} ]  , \label{eq:P2}\\
&X_{\{3\}}= \mathbb{G}_{2 \times 6} \ [f_{1,\{1,3\}};f_{2,\{1,3\}};f_{3,\{2,3\}};f_{4,\{2,3\}};f_{5,\emptyset};f_{6,\emptyset} ]  .\label{eq:P3}
\end{align}

We then consider $\Sc=\{1,2\}$, which contains two users. We have
\begin{align}
X_{\{1,2\}}= \mathbb{G}_{2 \times 6} \ [f_{1,\{2\}};f_{2,\{2\}};f_{3,\{1\}};f_{4,\{1\}};f_{5,\{1,2,3\}};f_{6,\{1,2,3\}} ] .\label{eq:P12}
\end{align}
From $X_{\{1,2\}}$, user $1$  caches all except $f_{1,\{2\}}$ and $f_{2,\{2\}}$, such that it can recover those two subfiles in $X_{\{1,2\}}$. In addition, user $2$ can recover $ f_{3,\{1\}}$ and $f_{4,\{1\}}$ from $X_{\{1,2\}}$.
Similarly, we have 
\begin{align}
&X_{\{1,3\}}=  \mathbb{G}_{2 \times 6}  \ [f_{1,\{3\}};f_{2,\{3\}};f_{3,\{1,2,3\}};f_{4,\{1,2,3\}};f_{5,\{1\}};f_{6,\{1\}} ] , \label{eq:P13}\\
&X_{\{2,3\}}=  \mathbb{G}_{2 \times 6} \ [f_{1,\{1,2,3\}};f_{2,\{1,2,3\}};f_{3,\{3\}};f_{4,\{3\}};f_{5,\{2\}};f_{6,\{2\}} ]  .\label{eq:P23}
\end{align}
 
 Finally we consider $\Sc=\{1,2,3\}$, which contains three users. We have 
\begin{align}
X_{\{1,2,3\}}= \mathbb{G}_{2 \times 6} \ [f_{1,\{2,3\}};f_{2,\{2,3\}};f_{3,\{1,3\}};f_{4,\{1,3\}};f_{5,\{2,3\}};f_{6,\{2,3\}} ].\label{eq:P123}
\end{align}
From $X_{\{1,2,3\}}$, user $1$  caches all except $f_{1,\{2,3\}}$ and $f_{2,\{2,3\}}$, such that it can recover those two subfiles in $X_{\{1,2,3\}}$. In addition, user $2$ can recover $f_{3,\{1,3\}}$ and $f_{4,\{1,3\}}$ while user $3$ can recover $f_{5,\{1,2\}}$ and $f_{6,\{1,2\}}$.

Hence, the server transmits $X=(X_{\Sc}: \Sc\subseteq [\Ksf], |\Sc|\in [3] )$, such that each user can recover its desired files in the delivery phase. For the privacy constraint in~\eqref{eq:privacy2}, we let then focus on the demand matrix  $\mathbb{D}=[1,2;1,3;1,4]$, and show the compositions of the received multicast messages by each user are  equivalent from its viewpoint to the ones for the demand matrix  $\mathbb{D}=[1,2;3,4;5,6]$.

{\it Delivery Phase for $\mathbb{D}=[1,2;1,3;1,4]$.}
 For  $\mathbb{D}=[1,2;1,3;1,4]$, we have $\Qc_{1}=\{1,2,3\}$, $\Qc_2=\{1\}$, $\Qc_3=\{2\}$, $\Qc_4=\{3\}$, and $ \Qc_5=\Qc_6=\emptyset$.
From the same way to construct multicast messages as described above, for  $\mathbb{D}=[1,2;1,3;1,4]$ we have 
\begin{align}
&X_{\{1\}}=  \mathbb{G}_{2 \times 6}  \ [f_{1,\{2,3\}};f_{2,\emptyset};f_{3,\{1,2\}};f_{4,\{1,3\}};f_{5,\{1\}};f_{6,\{1\}} ] , \label{eq:P1 d2}\\
&X_{\{2\}}=  \mathbb{G}_{2 \times 6} \ [f_{1,\{1,3\}};f_{2,\{1,2\}};f_{3,\emptyset};f_{4,\{2,3\}};f_{5,\{2\}};f_{6,\{2\}} ] , \label{eq:P2 d2}\\
&X_{\{3\}}=  \mathbb{G}_{2 \times 6} \ [f_{1,\{1,2\}};f_{2,\{1,3\}};f_{3,\{2,3\}};f_{4,\emptyset};f_{5,\{3\}};f_{6,\{3\}} ] , \label{eq:P3 d2}\\
&X_{\{1,2\}}=   \mathbb{G}_{2 \times 6} \ [f_{1,\{3\}};f_{2,\{2\}};f_{3,\{1\}};f_{4,\{1,2,3\}};f_{5,\{1,2\}};f_{6,\{1,2\}} ] , \label{eq:P12 d2}\\
&X_{\{1,3\}}=  \mathbb{G}_{2 \times 6} \ [f_{1,\{2\}};f_{2,\{3\}};f_{3,\{1,2,3\}};f_{4,\{1\}};f_{5,\{1,3\}};f_{6,\{1,3\}} ] , \label{eq:P13 d2}\\
&X_{\{2,3\}}=   \mathbb{G}_{2 \times 6} \ [f_{1,\{1\}};f_{2,\{1,2,3\}};f_{3,\{3\}};f_{4,\{2\}};f_{5,\{2,3\}};f_{6,\{2,3\}} ] , \label{eq:P23 d2}\\
&X_{\{1,2,3\}}=   \mathbb{G}_{2 \times 6} \ [f_{1,\emptyset};f_{2,\{2,3\}};f_{3,\{1,3\}};f_{4,\{1,2\}};f_{5,\{1,2,3\}};f_{6,\{1,2,3\}} ], \label{eq:P123 d2}
\end{align}
and let the server transmit   $X=(X_{\Sc}: \Sc\subseteq [\Ksf], |\Sc|\in [3] )$.

{\it Privacy.}
For any demand matrix (we do not list the transmission for all demand matrices for sake of simplicity), we can summarize four common points:
 \begin{enumerate}
\item  for any $i\in [\Nsf]$, each   subfile of $F_i$ cached by user $1$ is equivalent from the viewpoint of user $1$;   each   subfile of $F_i$ not cached by user $1$ is also equivalent from the viewpoint of user $1$;
\item there does not exist any subfile appearing in two multicast messages, which ensures both the decodability and privacy.
\item  in each of $X_{\{1\}}, X_{\{1,2\}}, X_{\{1,3\}}, X_{\{1,2,3\}}$, there is exactly one subfile of each file. If this subfile is from a file requested by user $1$, it is uncached by user $1$; otherwise, it is cached by user $1$. 
\item   in each of $X_{\{2\}}, X_{\{3\}}, X_{\{2,3\}}$,  there is exactly one subfile of each file. If this subfile is from a file requested by user $1$, it is cached by user $1$; otherwise, it is uncached by user $1$.   
\end{enumerate}
 Hence, from the viewpoint of user $1$,
 the composition  of $X$ (i.e., the subfiles in each XOR multicast message), 
is symmetric for different demand matrices  in which $\dv_1=(1,2)$. 
In other words, knowing $\dv_1$ and $Z_1$, the probability that  $X$ is generated for any demand matrix $\mathbb{D}_{\backslash\{1\}}$, is identical.
 Similarly, for any user in $[\Ksf]$, it cannot   get any information about the demands of other users neither. The formal information-theoretic proof on the privacy constraint in~\eqref{eq:privacy2} of the new private caching scheme can be found in Appendix~\ref{sec:privacy lemma 1}.

{\it Performance.}
For any demand matrix, we transmit $\binom{\Ksf}{1}+\dots+\binom{\Ksf}{\Ksf}=2^{\Ksf}-1=7$ multicast messages, each of which contains $\Lsf=2$ linear combinations of subfiles. Since each subfile has $\Bsf/8$ bits,  the load in the delivery phase is $14/8=1.75$ with sub-packetization level $8$. The achieved load by the virtual-user scheme in Theorem~\ref{thm:extended scheme} is $23/12\approx 1.92$ with sub-packetization level $ 2^{\binom{\Nsf}{\Lsf}\Ksf \Hc(\Msf/\Nsf)}\approx 2.47 \times 10^{13}.
$ Notice that the achieved load by the baseline scheme is $\Nsf-\Msf=3$. In conclusion, the achieved load by the MDS-based scheme is less than the virtual-user scheme, and with a much lower sub-packetization level.

\hfill$\square$
\end{example}

\begin{rem}
\label{rem:non overlap}
  Besides the high-level privacy strategy of the MDS-based scheme introduced in Section~\ref{sec:main results},
  there is another important construction   which makes the MDS-based scheme   private. 
  
In the multicast message $X_{\Sc}$, there is one subfile from each file. The subfile for the file $F_i$ is $f_{i, \Sc\cup\Qc_i \setminus (\Sc \cap \Qc_i)}$,  instead of $f_{i, \Sc \setminus \Qc_i}$, such that there does not exist any subfile appearing in two multicast messages.    We assume  $f_{i, \Sc \setminus \Qc_i}$ is transmitted in  $X_{\Sc}$. 
 For each demand matrix where   users request different  files, one subfile appears in at most two multicast messages, e.g., if
    $F_1$ is only demanded by user $1$,   $f_{1,\{3\}}$  appears in $X_{\{3\}}$ and $X_{\{1,3\}}$.
  However, if $F_1$ is demanded by both users $1,2$ and not by user $3$, it can be seen that $f_{1,\{3\}}$  appears in  $X_{\{1,3\}}$,  $X_{\{2,3\}}$, and $X_{\{1,2,3\}}$. Hence, the composition of the multicast messages depends on the users' demands. 
  \hfill$\square$
 \end{rem}

In the following example, we also consider $\Ksf=3$, $\Nsf=6$, $\Lsf=2$, but with $\Msf=24/7$ which leads $t=1$ in~\eqref{eq:scheme 2 1}. For $t\geq 1$, the new private caching scheme needs an MDS precoding in the placement phase.
\begin{example}[$\Ksf=3$, $\Nsf=6$, $\Msf=24/7$, $\Lsf=2$]
\label{exp:improved scheme2}
\rm
From~\eqref{eq:scheme 2 1}, we  can compute $t=1$.

{\it Placement Phase.}
Each file $F_i$ where $i\in [\Nsf]$ is divided into $2^{\Ksf-1}+\binom{\Ksf-1}{t}+\dots+\binom{\Ksf-1}{\Ksf-1}=7$ non-overlapping and equal-length pieces, which are then encoded by a $\left(2^{\Ksf},  2^{\Ksf-1}+\binom{\Ksf-1}{t}+\dots+\binom{\Ksf-1}{\Ksf-1} \right)=(8,7)$ MDS code (the parameters of the MDS code will be explained later).\footnote{\label{foot:t=0}When $t=0$, it can be seen that $2^{\Ksf-1}+\binom{\Ksf-1}{t}+\dots+\binom{\Ksf-1}{\Ksf-1} =2^{\Ksf}$. So we do not need  the MDS precoding.} Each MDS coded symbol has $\Bsf/7$ bits.
By the property of the MDS code, any $7$ MDS coded symbols can reconstruct the whole file. The $8$ MDS coded symbols of $F_i$ are denoted by $S^i_{1},\ldots,S^i_8$. The rest of the placement phase is the same as $t=0$ in  Example~\ref{exp:improved scheme}. More precisely, we  randomly  generate a permutation of $\left[2^{\Ksf} \right]$, denoted by $\pv_i=(p_{i,1},\ldots,p_{i,2^{\Ksf}})$ and
 assign each MDS coded symbol to a subfile according to  $\pv_i$ as in~\eqref{eq:example assignment}.
Each user $k\in [\Ksf]$ caches $f_{i,\Wc}$ if $k\in \Wc$. Hence, each user totally caches $\frac{4 \Bsf}{7}\Nsf=\frac{24 \Bsf}{7}=\Msf\Bsf$ bits satisfying the memory size constraint. 

{\it Delivery  Phase for $\mathbb{D}=[1,2;3,4;5,6]$.}
For each  subset $\Sc \subseteq [\Ksf]$ where $|\Sc|\geq t+1 =2$, we let the server  transmit $X_{\Sc}$  with the same construction in~\eqref{eq:P12}-\eqref{eq:P123}. 
 In other words, compared to Example~\ref{exp:improved scheme} with $t=0$, we only transmit $X_{\{1,2\}},X_{\{1,3\}},X_{\{2,3\}},X_{\{1,2,3\}}$. 
 
From   $X_{\{1,2\}}, X_{\{1,3\}}, X_{\{1,2,3\}}$, user $1$ can recover $3$ MDS coded symbols for each of its desired files. Since it caches $2^{\Ksf-1}=4$ MDS coded symbols for each file, it can recover each of its desired files by the  $4+3=7$ MDS coded symbols, and thus it can recover its desired files.

In short, the   $X_{\Sc}$'s where $0<|\Sc|<t+1$ are not transmitted in the delivery phase and thus each user cannot recover all subfiles of its desired files. Hence, we need the MDS precoding for $t\geq 1$. 

{\it Privacy.}
By the same reason as Example~\ref{exp:improved scheme}, the new private  scheme for $t=1$ can also satisfy the privacy constraint.

{\it Performance.}
For any demand matrix, we transmit $ 4$ multicast messages, each of which contains $\Lsf=2$ linear combinations of subfiles.  Since each subfile has $\Bsf/7$ bits, the load in the delivery phase is $8/7 \approx 1.14 $ with sub-packetization level $8$. The achieved load by the virtual-user scheme is $3550/2457 \approx 1.44$ with sub-packetization level $ 2^{\binom{\Nsf}{\Lsf}\Ksf \Hc(\Msf/\Nsf)}\approx 2.22 \times 10^{13}. $ Notice that the load achieved by the baseline scheme is $18/7 \approx 2.57$. As in Example~\ref{exp:improved scheme}, in this example the MDS-based scheme has a lower load and a much lower sub-packetzation level compared to the virtual-user scheme.  
\hfill$\square$
\end{example}

We are now ready to generalize Examples~\ref{exp:improved scheme} and~\ref{exp:improved scheme2}. We focus on the memory size 
$$
\Msf=\frac{2^{\Ksf-1}}{2^{\Ksf-1}+\binom{\Ksf-1}{t}+\binom{\Ksf-1}{t+1}+\dots+\binom{\Ksf-1}{\Ksf-1}}\Nsf,$$
 where $t\in [0:\Ksf-1]$. Notice that if $t=\Ksf$, we have $\Msf=\Nsf$ and each user can store the whole library in  its cache, such that the server needs not to transmit any packet in the delivery phase. 

{\it Placement Phase.}
Each file $F_i$ where $i\in [\Nsf]$ is divided into $2^{\Ksf-1}+\binom{\Ksf-1}{t}+\dots+\binom{\Ksf-1}{\Ksf-1}$ non-overlapping and equal-length pieces, which are then encoded by a $\left(2^{\Ksf},  2^{\Ksf-1}+\binom{\Ksf-1}{t}+\dots+\binom{\Ksf-1}{\Ksf-1} \right) $ MDS code. Each MDS coded symbol has $\frac{\Bsf}{2^{\Ksf-1}+\binom{\Ksf-1}{t}+\dots+\binom{\Ksf-1}{\Ksf-1}}$ bits, and the   MDS coded symbols of $F_i$ is denoted by $S^i_{1},\ldots,S^i_{2^{\Ksf}}$. 
We  randomly  generate a permutation of $\left[2^{\Ksf} \right]$, denoted by $\pv_i=(p_{i,1},\ldots,p_{i,2^{\Ksf}})$, independently and uniformly over the set of all possible permutations.
Recall that  $\text{Pow}(a,j)$ denotes the $j^{\text{th}}$ set in the power set of $[a]$ with a lexicographic order.
For each $j\in [2^{\Ksf}]$, we generate one subfile
\begin{align}
 f_{i,\text{Pow}(\Ksf,j)} := S^i_{p_{i,j}} .\label{eq:subfile assignment}
\end{align}
  Any $2^{\Ksf-1}+\binom{\Ksf-1}{t}+\dots+\binom{\Ksf-1}{\Ksf-1}$ subfiles of $F_i$ can reconstruct $F_i$.
For each $\Wc\subseteq [\Ksf]$, user $k\in [\Ksf]$ caches   $f_{i,\Wc}$ if $k\in \Wc$. It can be seen that each user caches $2^{\Ksf-1}$ subfiles of each file.
Hence, each user totally caches $\frac{2^{\Ksf-1}}{2^{\Ksf-1}+\binom{\Ksf-1}{t}+\dots+\binom{\Ksf-1}{\Ksf-1}}\Nsf\Bsf =\Msf \Bsf$ bits in its cache, satisfying the memory size constraint.

{\it Delivery Phase for $\mathbb{D}$.}
Recall that $\Qc_{i}$ where $i\in [\Nsf]$ denotes  the set of users demanding $F_i$.
For each  subset $\Sc \subseteq [\Ksf]$ where $|\Sc|\geq t+1$, 
the server generates
\begin{align}
X_{\Sc}= \mathbb{G}_{\Lsf \times \Nsf} \ [f_{1,\Sc\cup\Qc_1 \setminus (\Sc \cap \Qc_1)}; f_{2,\Sc\cup\Qc_2 \setminus (\Sc \cap \Qc_2)}; \dots; f_{\Nsf,\Sc\cup\Qc_{\Nsf} \setminus (\Sc \cap \Qc_{\Nsf})} ].\label{eq:general improved 1}
\end{align}
$X_{\Sc}$ contains $\Lsf$ linear combinations and 
in $X_{\Sc}$, each user $k\in \Sc$ caches all subfiles except $f_{i,\Sc\cup\Qc_i \setminus (\Sc \cap \Qc_i)}$ where $i\in \dv_k$. By the property of $ \mathbb{G}_{\Lsf \times \Nsf}$ (each $\Lsf$ columns are linearly independent), user $k $ can recover 
$f_{i,\Sc\cup\Qc_i \setminus (\Sc \cap \Qc_i)}$ where $i\in \dv_k$.
Then we let the server transmit
\begin{align}
X=(X_{\Sc}: \Sc \subseteq [\Ksf], |\Sc|\geq t+1).
\end{align}

{\it Decodability.}
 We first introduce the following lemma, which will be proved in Appendix~\ref{sec:deco lemma}.
\begin{lem}
\label{lem:deco}
For any demand matrix $\mathbb{D}\in \mathscr{D}$, there is no subfile   transmitted in more than one multicast message of the  scheme in Section~\ref{sub:improved 1}.
\end{lem}

We focus on  user $k\in [\Ksf]$ and file $F_i$ where $i\in \dv_k$. 
 For each  subset $\Sc \subseteq [\Ksf]$ where $|\Sc|\geq t+1$ and $k\in \Sc$, user $k$ can recover one uncached subfile of $F_i$ from  the multicast message $X_{\Sc}$. Considering all such subsets, user $k$ can recover $\binom{\Ksf-1}{t}+\dots+\binom{\Ksf-1}{\Ksf-1}$  uncached subfiles of $F_i$. By Lemma~\ref{lem:deco}, these subfiles are distinct. Hence, user $k$ can totally obtain $2^{\Ksf-1}+\binom{\Ksf-1}{t}+\dots+\binom{\Ksf-1}{\Ksf-1}$ subfiles of $F_i$ from the placement and delivery phases, such that it can recover $F_i$.

{\it Privacy.}
Let us focus on user $k$.
 Intuitively,  for each subfile of $F_i$ cached by  user $k$, since the random permutation $\pv_i$ is unknown to user $k$, it does not know the other users who also cache it, and thus each cached subfile of $F_i$ is equivalent from the viewpoint of user $k$. Similarly, each uncached subfile of $F_i$ is equivalent from the viewpoint of user $k$.
In each multicast message $X_{\Sc}$ where $\Sc\subseteq [\Ksf]$ and $|\Sc|\geq t+1$, 
\begin{itemize}
\item when $k \in \Sc$, there is exactly one subfile of each file. If this subfile is from a file requested by user $k$, it is uncached by user $k$; otherwise, it is cached by user $k$. 
\item when $k \notin \Sc$, there is exactly one subfile of each file. If this subfile is from a file requested by user $k$, it is cached by user $k$; otherwise, it is uncached by user $k$.  
\end{itemize}
 In addition,
by Lemma~\ref{lem:deco},    there does not exist any subfile  transmitted in more than one multicast messages. 
Hence,  the  compositions of  $X=(X_{\Sc}:k\in \Sc)$ for different demand matrices in which $\dv_k$ is the same, are equivalent from the viewpoint of user $k$.

In Appendix~\ref{sec:privacy lemma 1}, we will prove the privacy in a formal information-theoretic way.


{\it Performance.}
For any demand matrix, we transmit $\binom{\Ksf}{t+1}+\dots+\binom{\Ksf}{\Ksf}$ multicast messages, each of which contains $\Lsf$ linear combinations of subfiles. Since each subfile has  $\frac{\Bsf}{2^{\Ksf-1}+\binom{\Ksf-1}{t}+\dots+\binom{\Ksf-1}{\Ksf-1}}$  bits, the achieved load is $$\Lsf\frac{\binom{\Ksf}{t+1}+\dots+\binom{\Ksf}{\Ksf}}{2^{\Ksf-1}+\binom{\Ksf-1}{t}+\dots+\binom{\Ksf-1}{\Ksf-1}}=\Lsf\frac{2^{\Ksf}-\binom{\Ksf}{0}-\dots-\binom{\Ksf}{t} }{2^{\Ksf-1}+\binom{\Ksf-1}{t}+\dots+\binom{\Ksf-1}{\Ksf-1}},$$ as in~\eqref{eq:scheme 2 1}.
The sub-packetzation level is $2^{\Ksf}$.

\begin{rem}
\label{rem:generation to any placement}
It can be seen that in both of the above proposed schemes in Sections~\ref{sub:extended scheme} and~\ref{sub:improved 1},
 the   placement precoding which leads that  from the viewpoint of one user each cached subfile of one file is equivalent while each uncached subfile of one file  is also equivalent, is the key to preserve the privacy of the  demands of other users from this user. We refer this precoding as to {\it Private Placement Precoding}, which can be generalized as follows.

We focus on a caching placement with a  $(n,k)$ MDS precoding where $n\geq k$. Each file $F_i$ is divided into $k$ non-overlapping and equal-length pieces, which are then encoded by a  $(n,k)$ MDS code.  The   MDS coded symbols of $F_i$ is denoted by $S^i_{1},\ldots,S^i_{n}$, each of which contains $\Bsf/k$ bits.  We  randomly  generate a permutation of $[n]$, denoted by $\pv_i=(p_{i,1},\ldots,p_{i,n})$, independently and uniformly over the set of all possible permutations.
For each $j\in [n]$, we generate one subfile of each file $F_i$,
\begin{align}
f_{i,\Wc_j} := S^i_{p_{i,j}},\label{eq:general subfile assignment}
\end{align}
where $\Wc_j \subseteq [\Ksf]$ and we let each user in $\Wc_j$ cache $f_{i,\Wc_j}$. As a result, from the viewpoint of user $k$, each cached subfile of $F_i$ is equivalent from the viewpoint of user $k$, while each uncached subfile of $F_i$ is also equivalent. 
  It is obvious that when $n=k$, the placement is uncoded. Hence, the proposed private placement precoding can be also used with any uncoded cache placement. 

Even if we use the proposed   private precoding for the MAN coded caching scheme described in Section~\ref{sub:MAN}, the privacy constraint does not hold because the compositions of the MAN multicast messages are not symmetric for different demand matrices.
 \hfill $\square$ 
\end{rem}
\subsection{Proof of~\eqref{eq:scheme 2 2}}
\label{sub:improved 2}
When  $\Msf\geq \frac{2^{\Ksf-1}}{2^{\Ksf-1}+1}\Nsf$, by memory-sharing  between the corner points $t=\Ksf-1$ and $t=\Ksf$ in~\eqref{eq:scheme 2 1},  the MDS-based scheme in Section~\ref{sub:improved 2} achieves the load $\Lsf\left(1-\frac{\Msf}{\Nsf}\right)$, which coincides the converse bound for the MAN caching model with multiple request in~\cite{franarxiv}. 

In the following, we will introduce another private caching scheme for $\Msf=\frac{2\Ksf-1}{2\Ksf}\Nsf$. By memory-sharing between the corner points  $\Msf= \frac{2\Ksf-1}{2\Ksf}\Nsf$ and $\Msf_2= \Nsf$,  for any $\Msf_1 \geq \frac{2\Ksf-1}{2\Ksf}\Nsf$,   the load  $\Lsf\left(1-\frac{\Msf_1}{\Nsf}\right)$ is achievable. Hence, if $\frac{2\Ksf-1}{2\Ksf}\Nsf \leq \frac{2^{\Ksf-1}}{2^{\Ksf-1}+1}\Nsf$ (i.e., $\Ksf\geq 4$), we can replace the corner point in~\eqref{eq:scheme 2 1} with $t=\Ksf-1$ by the corner point  in~\eqref{eq:scheme 2 2}.

{\it Placement Phase.}
Each file $F_i$ where $i\in [\Nsf]$ is divided into $\binom{\Ksf}{\Ksf-1}+\Ksf\binom{\Ksf}{\Ksf}=2\Ksf$ non-overlapping and equal-length pieces,   denoted by $S^i_{1},\ldots,S^i_{2\Ksf}$, where each piece has $\frac{\Bsf}{2\Ksf}$ bits.
We  randomly  generate a permutation of $\left[2\Ksf\right]$, denoted by $\pv_i=(p_{i,1},\ldots,p_{i,2\Ksf})$, independently and uniformly over the set of all possible permutations.
For each $k\in [\Ksf]$, we generate one subfile   $f_{i,[\Ksf]\setminus\{k\}}=S^i_{p_{i,k}}$. In addition, for each $q\in [\Ksf]$, we also generate one subfile $f_{i,[\Ksf],q}=S^i_{p_{i,\Ksf+q}}$.

Each user $k\in [\Ksf]$ caches $f_{i,\Wc}$ where $\Wc\subseteq [\Ksf]$ and $|\Wc|=\Ksf-1$, if $k\in \Wc$. User $k$ also caches $f_{i,[\Ksf],q}$ for each $q\in [\Ksf]$.
Hence, each user totally caches $\frac{\binom{\Ksf-1}{\Ksf-2}+\Ksf\binom{\Ksf-1}{\Ksf-1} }{2\Ksf} \Nsf\Bsf =\Msf \Bsf$ bits in its cache, satisfying the memory size constraint.

{\it Delivery Phase for $\mathbb{D}$.}
Notice that each user caches $2\Ksf-1$ subfiles of each file, and thus it needs to recover one subfile of each of its desired files.

In the delivery phase, only one multicast message is generated and transmitted by the server,  
\begin{align}
X= \mathbb{G}_{\Lsf \times \Ksf\Nsf} \  \gv_{\mathbb{D}}.\label{eq:general improved 2}
\end{align}
$ \gv_{\mathbb{D}}$ is a vector containing $\Ksf\Nsf$ pieces. Define $\gv_{\mathbb{D}}(j)$ as the $j^{\text{th}}$ piece of $\gv_{\mathbb{D}}$, where $j\in [\Ksf\Nsf]$.
For each   $i\in [\Nsf]$ and each $k\in [\Ksf]$, 
\begin{itemize}
\item if $i\in \dv_k$ (i.e., user $k$ demands $F_i$), we let $\gv_{\mathbb{D}}\big( (i-1)\Ksf+k \big)= f_{i,[\Ksf]\setminus \{k\}}$;
\item otherwise, we let $\gv_{\mathbb{D}}\big( (i-1)\Ksf+k \big)= f_{i,[\Ksf],k}$.
\end{itemize}

{\it Decodability.}
Among the $\Ksf\Nsf$ subfiles in  $X$, each user $k\in [\Ksf]$ caches all except $f_{i,[\Ksf]\setminus \{k\}}$ where $i\in \dv_k$.  By the property of $ \mathbb{G}_{\Lsf \times \Ksf\Nsf}$ (each $\Lsf$ columns are linearly independent), user $k $ can recover these $\Lsf$ subfiles. Hence, we prove the decodability.

{\it Privacy.}
Let us focus on user $k$.
Intuitively, for any demand matrix,  there are exactly $\Ksf$ subfiles of each file in $P_{[\Ksf]}$. Among the $\Ksf$ subfiles of each file demanded by user $k$, user $k$ caches $\Ksf-1$ subfiles, while  among the $\Ksf$ subfiles of each file not demanded by user $k$, user $k$ caches $\Ksf$ subfiles.
In addition, for any  $i\in[\Nsf]$,  each  subfile of     $F_i$   cached by user $k$ is equivalent from the viewpoint of user $k$.
Hence, the   multicast message  $X$ for different demand matrices in which $\dv_k$ is the same, are equivalent from the viewpoint of user $k$. 

In Appendix~\ref{sec:privacy lemma 2}, we will prove the privacy in a formal information-theoretic way.

{\it Performance.}
For any demand matrix,  $P_{[\Ksf]}$ contains $\Lsf$ linear combinations of subfiles. Since each subfile has $\frac{\Bsf}{2\Ksf}$ bits, the achieved load is $\frac{\Lsf}{2\Ksf}$, as in~\eqref{eq:scheme 2 2}. The sub-packetzation level is $2\Ksf$.

\section{Conclusions}
\label{sec:conclusion}
In this paper, we introduced a novel shared-link caching model with private demands, while the objective is to design a two-phase caching scheme with minimum load while preserving the privacy of the users demands.  We believe that preserving the privacy of the users demands from other users that legitimately use the caching/content delivery system is an important problem that differs conceptionally from previously proposed models with eavesdroppers or private information retrieval (PIR), as shortly outlined in Section~\ref{sec:intro}.
For the formulated shared-link caching problem with private demands, we proposed   two novel private coded caching schemes, the virtual-user scheme and the MDS-based scheme, which are information-theoretically private. 
Compared to the existing converse bounds for the shared-link caching model without privacy constraint,  
the virtual-user scheme is order optimal within a constant factor when $\Nsf \leq \Lsf\Ksf$, or when $\Nsf<\Lsf\Ksf$ and $\Msf \geq \Nsf/\Ksf.$ In addition, both of the two schemes are order optimal within a factor of $2$ when $\Msf \geq \Nsf/2$.

The only open case where the multiplicative gaps between the proposed schemes and the existing converse bounds for the shared-link caching model without privacy constraint are not constant is when  $\Nsf<\Lsf\Ksf$ and $\Msf< \Nsf/\Ksf.$ In addition, since the virtual-user scheme has exponentially high sub-packetization level compared to the original MAN coded caching scheme and the 
MDS-based scheme does not have the order optimality results on the achieved load when $\Msf< \Nsf/2$, 
the problem of 
preserving the privacy of the demands in the regime $\Msf < \Nsf/2$ with order optimal load and small sub-packetization (at least not exponentially larger than the original MAN coded caching scheme remains open.
 On-going/future work  includes deriving a converse bound for this   caching model with privacy and designing improved private caching schemes  with small sub-packetization to solve the above two open problems.

 \appendices

 
\section{Proof of Order Optimality Results}
\label{sec:proof of order opitmality}
\subsection{Proof of Theorem~\ref{thm:extended order}}
\label{sub:proof of general order}
 

{\it Converse.} 
We use the existing converse bound  in~\cite{improvedlower2017Ghasemi,yas2,Sengupta2017multirequest} for the  shared-link caching model without privacy, which obviously provides a load lower bound for the  shared-link caching model with private demands. More precisely, we consider $\Lsf=1$ and $\Lsf>1$, respectively.
\begin{itemize}
\item $\Lsf=1$.  If  $\Nsf \leq \Ksf$,   the lower convex envelope of $(0,\Nsf)$ and $\left( \frac{\Nsf t^{\prime}}{\Ksf}, \frac{\Ksf-t^{\prime}}{t^{\prime}+1}\right)$ where $t^{\prime} \in [\Ksf]$ is order optimal within a factor of $4$~\cite{improvedlower2017Ghasemi}. If  $\Nsf>\Ksf$, the lower convex envelope of  $\left( \frac{\Nsf t^{\prime}}{\Ksf}, \frac{\Ksf-t^{\prime}}{t^{\prime}+1}\right)$ where $t^{\prime}\in [0:\Ksf]$ is order optimal within a factor of $2$~\cite{yas2} . In addition, as shown in~\cite{ontheoptimality} that the corner points  $\left( \frac{\Nsf t^{\prime}}{\Ksf}, \frac{\Ksf-t^{\prime}}{t^{\prime}+1}\right)$ where $t^{\prime}\in [0:\Ksf]$ are successively convex. Hence, when  $\Nsf>\Ksf$ and $\Msf \geq  \Nsf/\Ksf$, the lower convex envelop of $ \left( \frac{\Nsf t^{\prime} }{\Ksf }, \frac{\Ksf-t^{\prime}}{  t^{\prime}+1 } \right)$, where $t^{\prime}\in [\Ksf]$ is order optimal within a factor of $2$.

\item  $\Lsf>1$.  The lower convex envelope of $(0,\Nsf)$ and $\left( \frac{\Nsf t^{\prime}}{\Ksf}, \Lsf\frac{\Ksf-t^{\prime}}{t^{\prime}+1}\right)$ where $t^{\prime} \in [0:\Ksf]$ is order optimal within  a factor of $11$~\cite{Sengupta2017multirequest}.  If $\Nsf \leq \Lsf\Ksf$,  the lower convex envelope of $(0,\Nsf)$ and $\left( \frac{\Nsf t^{\prime}}{\Ksf}, \Lsf\frac{\Ksf-t^{\prime}}{t^{\prime}+1}\right)$ where $t^{\prime} \in [\Ksf]$ is order optimal within  a factor of $11$; if $\Nsf > \Lsf\Ksf$,  the lower convex envelop of  $\left( \frac{\Nsf t^{\prime}}{\Ksf}, \Lsf\frac{\Ksf-t^{\prime}}{t^{\prime}+1}\right)$ where $t^{\prime}\in [\Ksf]$, is order optimal within a factor of $2$ when $\Msf \geq  \Nsf/\Ksf$.
\end{itemize}

 {\it Achievability.}
We will prove that from the achieved corner points by the proposed scheme in Theorem~\ref{thm:extended scheme}, $\left(\frac{ \Nsf t }{\binom{\Nsf}{\Lsf} \Ksf}, \Lsf\frac{\binom{\Nsf}{\Lsf}\Ksf-t }{t +1} \right)$ where $t \in \left[ \binom{\Nsf}{\Lsf} \Ksf \right]$,
 we can achieve $\left( \frac{\Nsf t^{\prime} }{\Ksf }, 2\Lsf \frac{ \Ksf-t^{\prime} }{ t^{\prime}+1 } \right)$, where $t^{\prime} \in [\Ksf]$.  

We now focus on one $t^{\prime} \in [\Ksf]$. We let $t =\binom{\Nsf}{\Lsf} t^{\prime}$ and we can achieve
\begin{align}
\Rsf_{\text{v}}  &= \Lsf \frac{\binom{\Nsf}{\Lsf}\Ksf -t}{t+1} \nonumber\\
 &=\Lsf \frac{\binom{\Nsf}{\Lsf}\Ksf-  \binom{\Nsf}{\Lsf} t^{\prime} }{ \binom{\Nsf}{\Lsf} t^{\prime} +1}  \nonumber\\
 &=\Lsf \frac{\Ksf-t^{\prime}}{t^{\prime}+\frac{1}{\binom{\Nsf}{\Lsf}}}\nonumber\\
 &\leq  \frac{2(\Ksf-t^{\prime})}{t^{\prime}+1},\label{eq:order gen}
\end{align}
where~\eqref{eq:order gen} comes from 
$$\frac{t^{\prime}+1}{t^{\prime}+\frac{1}{\binom{\Nsf}{\Lsf}}} \leq \frac{t^{\prime}+1}{t^{\prime}}\leq 2, \ \text{when } t\geq 1.$$ 
Recall that $(0,\Nsf)$ can be also achieved by the proposed scheme. Hence, we prove Theorem~\ref{thm:extended order}.


 
 \subsection{Proof of Theorem~\ref{thm:general order}}
\label{sub:thm:general order}
 {\it Converse.} 
We use the existing converse bound  in~\cite{franarxiv} for the  shared-link caching model without privacy, which obviously provides a load lower bound for the  shared-link caching model with private demands.
From~\cite[Theorem 1]{franarxiv} with $s=1$, we have 
\begin{align}
\Rsf^{\star} \geq \Lsf \left(1-\frac{\Msf}{\Nsf} \right).\label{eq:converse order general}
\end{align}

 {\it Achievability.}
When $\Msf_1=\Nsf/2$, from~\eqref{eq:scheme 2 1} with $t=0$ achieved by the improved scheme, we have 
\begin{align}
\Rsf_{\text{m}}=\Lsf\frac{2^{\Ksf}-1}{2^{\Ksf}} \leq \Lsf.\label{eq:achievable order general} 
\end{align}
Hence, by memory-sharing between  $\Msf_1=\Nsf/2$ with load less than $\Lsf$ and $\Msf_2=\Nsf$ with load equal to $0$, we have for any $\Msf \in [\Nsf/2, \Nsf]$,
\begin{align}
\Rsf_{\text{m}} &\leq 2\Lsf\left(1-\frac{\Msf}{\Nsf} \right)\nonumber\\ 
& \leq 2\Rsf^{\star}, \label{eq:final order general}
\end{align}
 where~\eqref{eq:final order general} comes from~\eqref{eq:converse order general}.
 
 Similarly, by letting $t= \left\lfloor   \frac{\binom{\Nsf}{\Lsf} \Ksf}{2}  \right\rfloor $ in~\eqref{eq:extended scheme}, it can be proved that   when $\Msf_1=\Nsf/2$, $\Rsf_{\text{v}} \leq \Lsf$. Hence, $\Rsf_{\text{v}}$ is also order optimal within a factor of $2$ when $\Msf \geq \Nsf/2$.
 
 \section{Proof of Lemma~\ref{lem:deco}}
 \label{sec:deco lemma}
 It is equivalent to prove for any two sets $\Sc_1\subseteq [\Ksf]$ and $\Sc_2\subseteq [\Ksf]$ where $\Sc_1 \neq \Sc_2$, we have 
\begin{align}
(\Sc_1 \cup \Qc_i)\setminus (\Sc_1 \cap \Qc_i) \neq  (\Sc_2 \cup \Qc_i)\setminus (\Sc_2 \cap \Qc_i), \ \forall i\in [\Nsf].\label{eq:no equal}
\end{align} 
In addition,  for each $j\in [2]$, we let $\Sc_j=\Sc_{j,1}\cup \Sc_{j,2}$ where $\Sc_{j,1} \subseteq \Qc_i$ and $\Sc_{j,2}\cap \Qc_i =\emptyset$.

 Without loss of generality, we assume $|\Sc_1|\geq |\Sc_2|$. We focus on two cases:
 \begin{enumerate}
 \item $\Sc_{1,2} \neq \Sc_{2,2}$. It can be seen that   $\Sc_{j,2} \subseteq \left( (\Sc_j \cup \Qc_i)\setminus (\Sc_j \cap \Qc_i) \right)$ for each $j\in [2]$. Hence,~\eqref{eq:no equal} holds for this case.
 \item $\Sc_{1,2} = \Sc_{2,2}$ and $\Sc_{1,1} \neq \Sc_{2,1}$. Since $|\Sc_1|\geq |\Sc_2|$ and $\Sc_{1,1} \neq \Sc_{2,1}$, there exists at least one user in $\Qc_i$ (assume to be $k$) who is in $\Sc_{1,1}\setminus \Sc_{2,1}$. 
Hence, this user $k$ is in  $(\Sc_2 \cup \Qc_i)\setminus (\Sc_2 \cap \Qc_i)$ but not in $(\Sc_1 \cup \Qc_i)\setminus (\Sc_1 \cap \Qc_i)$. Hence,~\eqref{eq:no equal} holds for this case.
 \end{enumerate}
In conclusion, we prove   Lemma~\ref{lem:deco}.
 
\section{Proof of the Privacy for the New Scheme in~\eqref{eq:scheme 2 1}} 
 \label{sec:privacy lemma 1}
We consider $t=0$ in~\eqref{eq:scheme 2 1}. It can be seen when $t>0$, the transmitted multicast messages are included in the the transmitted multicast messages for $t=0$. Hence, if we prove the privacy for $t=0$, the privacy for $t>0$ can also be proved.

 For the new scheme in~\eqref{eq:scheme 2 1}, we want to prove the privacy constraint in~\eqref{eq:privacy2}, 
\begin{align}
I(\mathbb{D}_{\backslash\{k\}} ;  X | Z_k, \dv_k  )=0, \ \forall k\in[\Ksf].  
\end{align}
 We now focus on one user $k$, one demand vector $\dv_k$, and one cache realization $z_k$. Assume $( x_{\Sc}: \Sc \subseteq [\Ksf],   |\Sc|>0  )$ is a possible realization of $( X_{\Sc}: \Sc \subseteq [\Ksf],  |\Sc|>0   )$, given $\dv_k$ and $z_k$. We want to prove for any demand matrix $\mathbb{D}_{\backslash\{k\}}$, the probability 
 $$
 \Pr\{( X_{\Sc}: \Sc \subseteq [\Ksf],  |\Sc|>0  )=( x_{\Sc}: \Sc \subseteq [\Ksf],  |\Sc|>0  )| \dv_k, z_k,\mathbb{D}_{\backslash\{k\}} \}  
 $$ 
 does not depend on $\mathbb{D}_{\backslash\{k\}}$.
 
 For each $\Sc \subseteq [\Ksf]$ and each $i\in [\Nsf]$, we denote the coded MDS symbol of $F_i$ in $X_{\Sc}$ by $X_{\Sc,i}$. 
 We have 
 \begin{align}
& \Pr\{ ( X_{\Sc}: \Sc \subseteq [\Ksf], |\Sc|>0  ) =( x_{\Sc}: \Sc \subseteq [\Ksf], |\Sc|>0 )| \dv_k, z_k,\mathbb{D}_{\backslash\{k\}} \} \nonumber\\
&= \Pr\{  ( X_{\Sc,i}: \Sc \subseteq [\Ksf], |\Sc|>0  ,i\in [\Nsf])=( x_{\Sc,i}: \Sc \subseteq [\Ksf], |\Sc|>0 , i\in [\Nsf])| \dv_k, z_k,\mathbb{D}_{\backslash\{k\}} \} \nonumber\\
 &  = \prod_{i\in [\Nsf]} \Pr\{( X_{\Sc,i}: \Sc \subseteq [\Ksf],  |\Sc|>0  )=( x_{\Sc,i}: \Sc \subseteq [\Ksf],  |\Sc|>0   )| \dv_k, z_k,\mathbb{D}_{\backslash\{k\}} \},\label{eq:product all files}
 \end{align}
where~\eqref{eq:product all files} comes from that the placement permutations $\pv_{1},\ldots, \pv_{\Nsf}$ are independent. 
 
 We then focus on two cases:
\begin{itemize}
\item $i\in \dv_k$. It is claimed in Lemma~\ref{lem:deco}  that there does not exist any subfile appearing two multicast messages. Given $z_k$, there are $2^{\Ksf-1}$  MDS coded symbols of $F_i$ not in $z_k$, each of which should be in one different $X_{\Sc_1}$ where $\Sc_1 \subseteq [\Ksf]$ and $k\in \Sc_1$.
In addition, there are  $2^{\Ksf-1}$  MDS coded symbols of $F_i$   in $z_k$, each of which should be in one different  $X_{\Sc_2}$ where $\Sc_2 \subseteq [\Ksf]$, $k \notin \Sc_2$, and $|\Sc_2|>0$.  
 Hence, we have 
  \begin{subequations}
\begin{align}
&\Pr \Big\{( X_{\Sc,i}: \Sc \subseteq [\Ksf], |\Sc|>0  )=( x_{\Sc,i}: \Sc \subseteq [\Ksf], |\Sc|>0  ) \Big| \dv_k, z_k,\mathbb{D}_{\backslash\{k\}} \Big\} \\
&=\Pr\Big\{( X_{\Sc_1,i}: \Sc_1 \subseteq [\Ksf], k\in \Sc_1  )=( x_{\Sc_1,i}: \Sc_1 \subseteq [\Ksf], k\in \Sc_1  ), \nonumber\\& ( X_{\Sc_2,i}: \Sc_2 \subseteq [\Ksf], k\notin \Sc_2, |\Sc_2|>0  )=( x_{\Sc_2,i}: \Sc_2 \subseteq [\Ksf], k\notin \Sc_2, |\Sc_2|>0  )
\Big| \dv_k, z_k,\mathbb{D}_{\backslash\{k\}} \Big\} \\
&=\Pr\Big\{( X_{\Sc_1,i}: \Sc_1 \subseteq [\Ksf], k\in \Sc_1  )=( x_{\Sc_1,i}: \Sc_1 \subseteq [\Ksf], k\in \Sc_1  )
\Big| \dv_k, z_k,\mathbb{D}_{\backslash\{k\}} \Big\} \nonumber\\
&\Pr\Big\{( X_{\Sc_2,i}: \Sc_2 \subseteq [\Ksf], k\notin \Sc_2, |\Sc_2|>0  )=( x_{\Sc_2,i}: \Sc_2 \subseteq [\Ksf], k\notin \Sc_2, |\Sc_2|>0  )
\Big| \dv_k, z_k,\mathbb{D}_{\backslash\{k\}} \Big\} \label{eq:indep}\\
&=\left(\frac{1}{2^{\Ksf-1}!} \right)^2 , \label{eq:case in improved 1}
\end{align}
  \end{subequations}
where $!$ represents the factorial operation.~\eqref{eq:indep} comes from that each $ X_{\Sc_1,i}$ is not cached by user $k$ and each $ X_{\Sc_2,i}$ is cached by user $k$, and thus their realizations are independent given $z_k$ and $\mathbb{D}$.
\item $i\notin \dv_k$. It is claimed in Lemma~\ref{lem:deco}  that there does not exist any subfile appearing two multicast messages. Given $z_k$, there are $2^{\Ksf-1}$ coded MDS symbols of $F_i$   in $z_k$,  each of which should be in one different $X_{\Sc_1}$ where $\Sc \subseteq [\Ksf]$ and $k\in \Sc_1$.
In addition, there are $2^{\Ksf-1}$ coded MDS symbols of $F_i$   not in $z_k$,  each of which should be in one different $X_{\Sc_2}$ where $\Sc \subseteq [\Ksf]$, $k\notin \Sc_2$, and $|\Sc_2|>0$.
 Hence, from the same derivation as~\eqref{eq:case in improved 1} we have 
 \begin{align}
&\Pr\{( X_{\Sc,i}: \Sc \subseteq [\Ksf],|\Sc|>0 )=( x_{\Sc,i}: \Sc \subseteq [\Ksf], |\Sc|>0)| \dv_k, z_k,\mathbb{D}_{\backslash\{k\}} \} =\left(\frac{1}{2^{\Ksf-1}!} \right)^2 . \label{eq:case in improved 2}
\end{align}
 \end{itemize} 
It can be seen   both of the probabilities in~\eqref{eq:case in improved 1} and~\eqref{eq:case in improved 2} are independent of $\mathbb{D}_{\backslash\{k\}}$. Hence, we can prove the probability in~\eqref{eq:product all files} is also independent of  $\mathbb{D}_{\backslash\{k\}}$. In conclusion, we prove the privacy constraint in~\eqref{eq:privacy2}.
 
 \section{Proof of the Privacy for the New Scheme in~\eqref{eq:scheme 2 2}} 
 \label{sec:privacy lemma 2}
For the new scheme  in~\eqref{eq:scheme 2 2}, there is only one multicast message in  $X$. 
We   want to prove the privacy constraint in~\eqref{eq:privacy2}, 
\begin{align}
I(\mathbb{D}_{\backslash\{k\}} ; X | Z_k, \dv_k  )=0, \ \forall k\in[\Ksf].  
\end{align}
  We also  focus on one user $k$, one demand vector $\dv_k$, and one cache realization $z_k$. Assume $x $ is a possible realization of $ X $, given $\dv_k$ and $z_k$. We want to prove for any demand matrix $\mathbb{D}_{\backslash\{k\}}$, the probability 
 $$
 \Pr\{ X =x | \dv_k, z_k,\mathbb{D}_{\backslash\{k\}} \}  
 $$ 
 does not depend on $\mathbb{D}_{\backslash\{k\}}$.
 
In $X $, there are $\Ksf$ pieces of each file $F_i$. Hence, $X_{ i}$ now denotes the set of $\Ksf$ pieces of $F_i$ in $X $. Since  the placement permutations $\pv_{1},\ldots, \pv_{\Nsf}$ are independent,
we have 
 \begin{align}
&  \Pr\{ X =x | \dv_k, z_k,\mathbb{D}_{\backslash\{k\}} \}  = \prod_{i\in [\Nsf]} \Pr\{ X_{ i}=x_{ i}| \dv_k, z_k,\mathbb{D}_{\backslash\{k\}} \}.\label{eq:product all files 2}
 \end{align}
  We also focus two cases:
\begin{itemize}
\item $i\in \dv_k$. Notice that $z_k$   contains $2\Ksf-1$ pieces of $F_i$ while $F_i$ contains $2\Ksf $ pieces. In addition, in $X_{ i}$ there are $\Ksf-1$ pieces   of $F_i$ cached in $z_k$ and one piece of $F_i$ not cached in $z_k$. Hence, we have 
\begin{align}
&\Pr\{ X_{ i}=x_{ i}| \dv_k, z_k,\mathbb{D}_{\backslash\{k\}} \} =\frac{1}{\binom{2\Ksf-1}{\Ksf-1}} . \label{eq:case in improved 12}
\end{align}
\item $i\notin \dv_k$. In $X_{ i}$ there are $\Ksf $ pieces   of $F_i$ cached in $z_k$. Hence, we have 
\begin{align}
&\Pr\{ X_{ i}=x_{ i}| \dv_k, z_k,\mathbb{D}_{\backslash\{k\}} \}=\frac{1}{\binom{2\Ksf-1}{\Ksf }}=\frac{1}{\binom{2\Ksf-1}{\Ksf-1}}.  \label{eq:case in improved 22}
\end{align}
\end{itemize} 
It can be seen   both of the probabilities in~\eqref{eq:case in improved 12} and~\eqref{eq:case in improved 22} are independent of $\mathbb{D}_{\backslash\{k\}}$. Hence, we can prove the probability in~\eqref{eq:product all files 2} is also independent of  $\mathbb{D}_{\backslash\{k\}}$. In conclusion, we prove the privacy constraint in~\eqref{eq:privacy2}.

\bibliographystyle{IEEEtran}
\bibliography{IEEEabrv,IEEEexample}


\end{document}